\documentstyle[emulateapj]{article}
\input epsf.tex
\newcommand{\be}{\begin{equation} }
\newcommand{\ee}{\end{equation}}
\begin{document}
\title{Cooling models for old White Dwarfs}
\author{Brad M. S. Hansen }
\affil{Canadian Institute for Theoretical Astrophysics, University of
Toronto, Toronto, ON M5S 3H8, Canada}
\authoremail{hansen@cita.utoronto.ca}

\begin{abstract}
We present new white dwarf cooling models which incorporate an accurate outer boundary
condition based on new opacity and detailed radiative transfer calculations. We find that helium atmosphere
dwarfs cool considerably faster than has previously been claimed, while old hydrogen atmosphere
dwarfs will deviate significantly from black body appearance. We use our new models
to derive age limits for the Galactic disk. We find that the 
 Liebert, Dahn \& Monet (1988)
luminosity function yields an age of only $6$~Gyr if it is complete to stated limits. However, age
estimates of individual dwarfs and the luminosity function of Oswalt et al (1995) are both
consistent with disk ages as large as $\sim$11~Gyr. We have also used our models to place constraints on white dwarf
dark matter in Galactic halos. We find that previous attempts using inadequate cooling models
were too severe and that direct detection limits allow a halo that is $11$~Gyr old. If the halo
is composed solely of helium atmosphere dwarfs, the lower age limit is only $7.5$~Gyr. We also
demonstrate the importance of studying the cooling sequences of white dwarfs in Globular clusters.
\end{abstract}


\section{Introduction}

The usefulness of white dwarf stars as stellar chronometers was recognised
many years ago (Schmidt 1959), but only in the last decade have the observational
data and theoretical models reached the level of sophistication necessary to
provide meaningful constraints on their parent populations. The determination
of the luminosity function of white dwarfs in the Galactic disk (Liebert et al 1979;
Liebert, Dahn \& Monet 1988) showed the presence of a turnover at faint magnitudes.
The interpretation of this turnover as being a consequence of the finite Galactic age  has been
used as a constraint on the history of local star formation (Winget et al 1987; Wood 1992;
Hernanz et al 1994) as have studies of the luminosity function of white dwarfs in
common proper motion binaries (Oswalt et al 1996).

Tamanaha et al (1990) used the Liebert, Dahn \& Monet (hereafter LDM) proper motion sample
 to
constrain the number density of white dwarfs in the Galactic halo. Motivation for this
work was provided by the suggestion (Larson 1986; Silk 1991) that some or all of the
dark matter in Galactic halos may be in the form of white dwarfs. This suggestion
has recently received a fresh impetus (Adams \& Laughlin 1996; Chabrier, Segretain \& Mera 1996;
Graff, Laughlin \& Freese 1998) from the results of studies of gravitational
microlensing towards the Large Magellanic Cloud (Paczynski 1986; Alcock et al 1997; Aubourg 1995). The use of
white dwarfs to constrain the ages of open (Von Hippel, Gilmore \& Jones 1995; Richer et al 1998) and globular
clusters (Renzini et al 1996; Richer et al 1997) has also been pursued with some success.

In addition to `traditional' white dwarfs, the peculiar class of helium core dwarfs
which result from binary evolution have received increasing observational (Kulkarni 1986;
Van Kerkwijk 1996;
Landsman et al 1997) and theoretical (Vennes, Fontaine \& Brassard 1993; Benvenuto \& Althaus 1998;
Hansen \& Phinney 1998a,b) attention. In particular, they have been used to constrain the ages and properties
of
millisecond pulsars (Kulkarni 1986; Hansen \& Phinney 1998b; Hansen 1998a).

The usefulness of old white dwarfs is not restricted simply to age determination. The cosmologically
relevant timescales for white dwarf evolution provide a testing ground for such esoterica as the
variation of fundamental constants (Garcia-Berro et al 1995) and the
 flux of magnetic monopoles (Freese 1984; Freese \& Krasteva 1998).

Given the above list of interesting applications, it becomes important to examine the accuracy of
the white dwarf models used in these analyses. In particular, the recent advances in the treatment
of white dwarf atmospheres (Bergeron, Wesemael \& Fontaine 1991; Bergeron, Saumon \& Wesemael 1995, hereafter BSW; 
Bergeron, Wesemael \& Beauchamp 1995)
suggest the need for an
 improvement in the treatment of the outer boundary
conditions in the cooling models, previously based on the grey atmosphere approximation. The cooling of
old white dwarfs is very sensitive to the outer boundary condition (e.g. D'Antona \& Mazzitelli 1990)
and so this is an important concern.
 The primary aim
of this paper is to present a detailed set of cooling models based on a proper radiative transfer
treatment of the outer boundary condition and to consider the astrophysical implications thereof.
In addition, these models represent the first comprehensive set of models to cover the full mass
range encompassing both carbon/oxygen and helium core white dwarfs appropriate for cosmochronological purposes.

In section~\ref{Outer} we discuss our atmospheric treatment and the consequences for white dwarf
cooling. In section~\ref{Cryst} we describe our treatment of the physics in the white dwarf
core (in particular crystallisation). Section~\ref{Results} describes a comprehensive set of
cooling curves appropriate for the determination of stellar ages and section~\ref{App}
discusses the consequences for several of the astrophysical problems outlined above.

\section{Atmospheric Treatment}
\label{Outer}

The importance of the atmospheric treatment stems from the strongly constrained
thermal profile of an old white dwarf. In the degenerate core the energy transfer
is dominated by electron conduction, a very efficient mechanism which keeps the
core essentially isothermal. In the outer parts convection dominates, so that the
temperature profile is determined by the equation of state. The convection extends
to the photosphere (B\"{o}hm et al 1977) so that there is no `radiative buffer' where
adjustments may occur to compensate for changes in atmospheric parameters and thereby
keep the core temperature unaffected. Similar considerations arise in giant planet
studies (Guillot et al 1995).
 Thus, changes in atmospheric parameters are
reflected directly in changes in core temperature and, since white dwarf cooling
is driven largely by the slow leakage of the thermal reservoir stored in the core,
directly in the cooling rate.

Over the last decade, most white dwarf cooling calculations have used (entirely or in
part) the
results of Wood (1992; 1995), an offspring of the code of Lamb \& Van Horn (1975).
The atmospheric treatment used there is based on grey atmospheres and Rosseland mean opacities,
using results from the OPAL group (Rogers \& Iglesias 1992) where applicable and, for cooler models,
the tabulated results of Lenzuni, Chernoff \& Salpeter (1991). The accuracy of the grey
approximation is determined by the photospheric opacity. If the opacity is approximately
constant over the appropriate wavelength range, then it may be well described by a mean
opacity (either Rosseland or Planck) and the emergent spectrum will resemble a black body.
Once the opacity becomes strongly peaked in some wavelength region, the approximation fails
and a proper radiative transfer calculation is required. Another independent cooling code (Benvenuto \& Althaus 1999)
has recently appeared, but suffers from the same lack of accurate opacities and radiative transfer at low
 temperatures\footnote{Benvenuto \& Althaus have also criticised the mass-radius relation of Hansen \& Phinney (1998a). Although
the level of disagreement is immaterial for the cooling evolution, we note that the mass-radius relations of the models
presented 
here agree with those of Wood at the 1\% level.}.

A simple test of the grey atmosphere approximation is to compare the values of the Rosseland and
Planck means (see e.g. Mihalas 1970 or Rybicki \& Lightman 1979 for definitions). When the opacity
$\kappa$ is approximately constant with wavelength, the two means are similar
in magnitude. If they deviate significantly then a proper radiative transfer treatment is required.
Figure~\ref{RossPlanck} shows the comparison between the two mean opacities as a 
function of temperature at fixed density ($\rho = 10^{-9} g/cm^3$) for a pure hydrogen atmosphere. The deviation
for $T < 5000$~K is a consequence of the formation of molecular hydrogen.

For the effective temperatures appropriate for the oldest white dwarfs ($T_{\rm eff} < 5000$~K), the
atmospheric constituents are neutral. The large gravities and associated pressure and temperature gradients
 of white dwarfs lead to rapid
separation of elements in the atmosphere, so that the atmospheric constituents are either
hydrogen or helium. The distribution of helium and hydrogen layer masses in old white
dwarfs is a subject with a long and controversial history (see e.g. Kepler \&
Bradley 1995) and one of the prime motivations behind the field of white dwarf
asteroseismology (Nather 1993). For our purposes, the processes that determine the
chemical composition are largely  immaterial, as the
changes in the relative fractions of DA and non-DA stars (in terminology of Sion et al 1983) occur at temperatures 
$>10^4$K\footnote{We will show later that the `non-DA gap' near $5000$K identified by
Bergeron, Ruiz \& Leggett (1995) is a simple consequence of the different cooling rates
of the two populations},
whereas white dwarfs spend most of their cooling ages below this temperature. The relative
proportions of hydrogen and helium atmosphere dwarfs are equal for cool white dwarfs (although
Bergeron, Ruiz \& Leggett 1998 suggest that the hydrogen fraction may be higher than previously
thought). The actual values of the hydrogen and helium layers are important however, as is
the possibility of trace constituents.

Asteroseismological observations of DA stars suggest that many have `thick' hydrogen envelopes,
 i.e. mass fractions $\sim 10^{-4}$, (Bradley 1998) and the results are consistent with all
DA stars having similar structures (Clemens 1993) although not conclusively so as yet. This value
is that expected from stellar evolution calculations (Iben 1984; Sch\"{o}nberner 1983; D'Antona \& Mazzitelli 1979;
Wood \& Faulkner 1986). The mass fractions of helium layers are in the range $10^{-6}-10^{-2}$
based on asteroseismological (Bradley \& Winget 1994; Nitta \& Winget 1998) or chemical dredge-up
(Pelletier 1986; Macdonald, Hernanz \& Jose 1998) considerations. The composition of the atmospheres inferred
from spectroscopic comparisons (BSW; Bergeron, Wesemael \& Beauchamp 1995) suggests that a small admixture
of helium in hydrogen atmospheres is possible. We shall see that hydrogen atmospheres are insensitive to
small admixtures of helium, but that helium atmospheres are very sensitive to small admixtures of
hydrogen (essentially, a small amount of hydrogen in a helium atmosphere increases the opacity
dramatically, while the converse is not true).

\subsection{Radiative Transfer}

The formation of molecular hydrogen leads to absorption in wide bands centred on the
collisionally induced roto-vibrational transitions of the H$_2$ molecule for $\lambda > 1 \mu$m (for
a general review see Borysow \& Jorgensen 1999).
The result of such strong infra-red absorption is to drive much of the emergent flux blueward,
causing large deviations from black-body appearance and significant departures from
grey atmosphere behaviour (see Figure~\ref{Photop}). Detailed calculations of these effects for white dwarf atmospheres
have been performed by Bergeron and collaborators (Bergeron et al 1991; BSW) and their
effects on observational appearance demonstrated. However, the cooling calculations have
not kept pace with these developments and so the subsequent attempts to derive cooling ages
(Bergeron, Ruiz \& Leggett 1997, hereafter BRL; Leggett, Ruiz \& Bergeron 1998, hereafter LRB) are not completely self-consistent.

To properly treat the cooling behaviour of white dwarfs with non-grey atmospheres, we have
performed our own set of radiative transfer calculations. As in the case of our calculation
of mean opacities (see Hansen \& Phinney 1998a) we have used the opacity microphysics from
Lenzuni et al (1991) with additional collisionally induced absorption cross-sections from
Zhang \& Borysow (1995). Using these opacities, we solve the radiative transfer equation
using the Feautrier and Avrett-Krook methods (e.g. Mihalas 1970) and including a mixing
length prescription for convection. The numerical treatment of convection can produce
convergence problems (BSW;  Saumon et al 1995) caused by a local
minimum in the opacity (as a function of temperature at fixed density) near T$ \sim 5000$~K.
This occurs where the dominant opacity changes from that due to $\rm H^-$ to that due to $\rm H_2$ molecular
absorption. This local minimum can cause discontinuous jumps in temperature as a function
of optical depth. However, we note from the results of BSW that such discontinuities occur
at low optical depth, where the temperature profile is essentially isothermal. Thus, to
avoid these convergence problems, we enforce isothermal atmospheres below optical depths
$\tau < 0.05$ or above the convection region (if the convective region extends to optical
depths that small). This means that the outer part of the grid acts as a single point
in the temperature correction procedure but as a full grid when calculating the flux
through the atmosphere. Comparison with the results of BSW indicates that this is a
robust procedure. For the purposes of comparison between different atmospheres we
use the Rosseland photosphere, i.e. the position in the atmosphere where the Rosseland
mean opacity is 2/3.

For a pure hydrogen atmosphere, the photospheric density increases as the white dwarf
cools. This trend continues down to T$_{\rm eff} \sim 3000$~K whereafter the photosphere
moves again to smaller densities. This occurs because, at these temperatures, the
Planck function peaks at precisely those wavelengths where the $\rm H_2$ absorption is
strongest. The resultant increase in the Rosseland mean opacity means a smaller
column is required and hence a lower photospheric density.
 For pure helium atmospheres, molecules do not form upon recombination,
 so that
the photospheric density continues to increase. This trend is only halted when the
densities get high enough that pressure ionization becomes important as a source of
electrons and hence opacity. As can be seen from Figure~\ref{Phot}, the photospheric
densities of hydrogen and helium atmospheres differ dramatically, a consequence of
the differences in opacity, as shown in Figure~\ref{Photop}.

These results provide the outer boundary condition for the cooling models in the form
of tables of temperature and pressure at a specified point in the atmosphere as a
function of luminosity and radius (or effective temperature). The exact location of
the point at which the condition is enforced is not very important, usually taken
to be the point at which the Rosseland opacity $\tau_{\rm R}=100$.  Thus, the variations
in density resulting from atmospheric composition translate directly into the marked
differences in cooling shown in the subsequent sections.

The radiative transfer calculations also provide the emergent spectrum of radiation
from these cool dwarfs. As the hydrogen atmospheres cool below 5000~K they show 
progressively larger deviations from black body appearance (Figure~\ref{Bands}) . Helium atmospheres don't
show the same dramatic differences but do show some deviations resulting from the
fact that a major opacity source at short wavelengths in these atmospheres is Rayleigh scattering from
neutral helium atoms (see Figure~\ref{Photop}). Scattering, as opposed to true absorption, simply redistributes
radiation in angle, rather than thermalising it, leading to deviations from black
body appearances in atmospheres with significant scattering contributions (e.g. Mihalas 1970).

The deep convection zones which extend all the way to the photosphere raise the
possibility of mixed hydrogen and helium compositions in the atmosphere. As expected,
the photospheres lie somewhere in between the pure hydrogen and helium mixtures.
The observational appearance, however, deviates quite significantly from either (as noted in BRL),
as shown in Figure~\ref{Bands}.
The strongly wavelength-dependent molecular absorption is strongly affected by the higher
densities in atmospheres with reduced amounts of hydrogen, so that mixed hydrogen/helium
atmospheres deviate more from a black body than pure hydrogen atmospheres. Based on these considerations,
BRL find that the vast majority of cool white dwarfs are consistent with pure hydrogen or helium
compositions.

At this point it is worth noting what of the above analysis we may consider on solid ground
and what is still subject to some uncertainty. The primary purpose of this calculation is to establish
the true boundary condition applicable to old white dwarfs whose atmospheres deviate strongly
from the grey approximation. For this purpose the results presented here may be considered robust (as confirmed
recently by by Saumon \& Jacobsen 1999),
because they essentially rest on establishing the appropriate photospheric densities and
temperatures. Even in the most uncertain case, namely that of very cool helium atmospheres whose
atmospheres are dominated by pressure ionisation, the results are robust, because pressure
ionisation only occurs in a limited density range near $\rho \sim 100 g/cm^3$. Note also that
these results are not sensitive to trace amounts of hydrogen (helium) in otherwise pure helium (hydrogen)
atmospheres, because the amounts allowed by the model atmospheres (BRL) are not enough to alter
the photospheric densities much.

We will also present detailed calculations of the optical and infra-red appearance of these
objects, in a similar fashion to the calculations of BSW and BRL. Note
that these results are not intended to supplant the  previous atmosphere calculations but are
rather presented to provide a comprehensive and self-consistent description of the white dwarf
cooling models presented below. In particular, we have not treated line formation and broadening in these atmospheres and thus the
determination of individual white dwarf parameters is not possible to the accuracy obtained by
BRL. However, we have extended our results to lower temperatures than BSW.
The lower temperature limit of 4000~K used by BSW was due to the looming spectre of hydrogen
pressure ionisation at lower temperatures. However, we have noted that the photosphere moves to
lower density again below 3000~K because the black body peak is located in regions of high molecular 
opacity (see above).

The calculations of observational appearance are somewhat less certain than the cooling calculations
themselves. The general character of the solutions is robust but residual uncertainties regarding
convection, conduction and pressure ionisation do introduce some uncertainty.
 Bergeron, Wesemael \& Fontaine (1991) find that atmospheric calculations are insensitive to different parameterisations
of mixing length convection. The influence of electron conduction (Kapranidis 1983) is small
(BSW) but may become increasingly important at lower temperatures, as are
the contributions to the opacity of transitions to higher rotovibrational states in the hydrogen molecules
(Zhang \& Borysow 1995).

\section{Crystallisation and Quantum Corrections}
\label{Cryst}

The application of these new boundary conditions represents the primary purpose of this
paper. However, some issues regarding the treatment of the core physics also require
attention in order to make comparisons with extant models in the literature.

Our code was originally developed to describe low mass helium
core white dwarfs and the basic details and tests of the code can be found in Hansen \& Phinney 1998a.
The Coulomb interactions in a helium core are not strong enough to cause crystallisation
so no detailed description of this process was included in the original code. However,
for stars with carbon and oxygen interiors, crystallisation is an important factor in the
evolution and below I describe it's implementation in the cooling code.

Crystallisation is important for two reasons. The first is that it provides a source of
extra energy. Apart from the release of $\sim{\rm k T}$ per ion of
 latent heat of crystallisation there may also be an additional energy release associated
with the chemical fractionation between solid and liquid phases in a mixed fluid
(Stevenson 1980; Garcia-Berro et al 1988; Segretain \& Chabrier 1993). Original estimates
of the importance of minor species such as $^{22} \rm Ne$ ( Segretain et al 1994; Hernanz et al 1994) have been
reduced with further calculation (Segretain 1996) but the fractionation of the primary
constituents  carbon and oxygen may prove important, introducing delays of $\sim 1-2$~Gyr
in the cooling of the faintest white dwarfs. The exact contribution (or even if  there is
any at all) is still debatable, however. This is because the continued operation of
the separation requires that the oxygen-depleted liquid region remain well mixed, possibly by
Rayleigh-Taylor instabilities (Mochkovitch 1983; Isern et al 1997). Furthermore, the
amount of energy released will depend on the chemical stratification of material in
the proto-white dwarf core during previous stellar evolution stages, which is itself
somewhat uncertain because of uncertainties in the cross-section for the
$^{12}C(\alpha,\gamma)^{16}O$ nuclear burning cross-section (see Salaris et al 1997 and
references therein). Thus, we shall compare cooling models with and without the
chemical separation energy contribution. It will also be shown that the importance
of the above effects is dependent somewhat on the aforementioned boundary conditions, because
faster cooling lessens the effect of a given energy release.

The second important effect of core crystallisation is the change in the heat capacity due
to the formation of a Coulomb lattice. As the star cools further, the heat capacity drops
rapidly as the crystal enters
the Debye regime where fewer of the normal modes of the lattice are excited.
 The influence
of quantum effects on the heat capacity of the liquid state (Chabrier, Ashcroft \& De Witt 1992)
are also included, although, as shown by Chabrier (1993), this correction only affects the
more massive white dwarfs.

To implement crystallisation (in particular the release of latent heat)
 with a Henyey code (which converges poorly when the physical
inputs are discontinuous), we release the latent heat between $\Gamma=165$ and $\Gamma=185$, i.e. we
consider crystallisation to occur at $\Gamma=\Gamma_c=175$ (Slattery, Doolen \& De Witt 1982; 
Farouki \& Hamaguchi 1993) where $\Gamma$ is the familiar Coulomb coupling parameter.
 Since the separation energy is also
released upon crystallisation we may consider it as an extra latent heat, albeit a function
of local composition. The energy per ion released is taken to be (e.g. Chabrier 1998)
\be
\frac{\Delta u}{kT} = - 0.9 \Gamma_e \Delta X \left[ \frac{Z_1^{5/3}}{A_1} - \frac{Z_2^{5/3}}{A_2} \right]
\ee
where $\Gamma_e = \Gamma / <Z>^2$ and $\Delta X$ is the difference in composition between
the newly crystalline material and the instantaneous ionic liquid composition (which contain ions of
mass and charge $A_{1,2}$ and $Z_{1,2}$ respectively). Dividing this
by the latent heat yields an `enhancement factor'
\be
q = 1 + 1.526 \left( \frac{ \Delta X}{ (<Z>/6)^2 } \right) \left( \frac{\Gamma_c}{175} \right)
\ee
where $<Z>$ is the mean charge in the pre-crystallisation liquid. Thus, the energy released is approximately
$30 \%$ more than that released by pure latent heat. We have used the crystallisation curves of
Segretain \& Chabrier (1993) to calculate the crystallisation of C/O mixtures assuming
uniform mixing in the liquid core region. Although the energy is, in fact, released over much of
the core (since the origin is the gravitational binding energy released in the rearrangement of the
density profile), approximating it as a localised extra latent heat will not provide a significant
source of error because of the efficient heat transport properties of the degenerate, isothermal core.

\section{Results}

\label{Results}

Using the above modifications to the code of HP, we have calculated a series of
cooling models appropriate to the study of faint white dwarfs. In addition to the
major modifications mentioned above, we have included hydrogen burning by the pp-process
for the hot models with hydrogen envelopes. This is important for the more massive
hydrogen envelope white dwarfs, because it sets an upper limit on the mass of hydrogen
allowed on the surface of the white dwarf (an effect not included in Wood's 
models)\footnote{This limit comes from steady burning only. The limits could be even more stringent
if the self-induced novae of Iben \& Macdonald (1986) occur}.
This could be important if some of the chemically peculiar white dwarfs owe their
compositional idiosyncrasies to dredge-up by deep convection zones. Figure~\ref{masses}
shows the magnitude of this effect.
We have also taken account of whether the convection zone gets deep enough to dredge
up material directly from deeper layers of different composition. This can serve to
limit the allowed thicknesses of helium layers which show small enhancements of
carbon.

\subsection{Comparison with other models}
The models of Wood (1992, 1995) are the standard that has been adopted over the last
decade for the study of cool white dwarfs. The envelope $L-T_c$ relations from these
models are also used in the series of cooling curves published by the European
group beginning with Hernanz et al (1994) and culminating in the paper by Salaris et al (1997) (hereafter SDG).
Thus, these models serve as a convenient template with which to analyse the changes introduced
by our modifications.

Figure~\ref{LH} shows a comparison of our models and Wood's for a standard DA model. The envelope
consists of a hydrogen layer of mass fraction $q_{\rm H} = 10^{-4}$ atop a helium layer of mass fraction
$q_{\rm He} = 10^{-2}$. We have calculated models for $0.6 \rm  M_{\odot}$ and core compositions of pure carbon, pure oxygen
and mixed C/O according to the appropriate profile of SDG. These are compared with
the corresponding pure carbon and pure oxygen models of Wood (1995).
The inclusion of proper outer boundary conditions leads to a
significant ($\sim 2$~Gyr) increase in the cooling ages at low luminosities. The mixed C/O model
lies midway between the C and O curves. It should be remembered, however, that this model is
$\sim 90 \%$ oxygen in the centre after crystallisation.

Comparison in the helium atmosphere case is more difficult. The standard Wood model of this
type has a helium mass fraction
$q_{\rm He} = 10^{-4}$. This was consistent with the determinations of helium envelope mass by
Pelletier et al (1986), who studied the dredge up of carbon from a diffusion tail to make
DQ white dwarfs. Recently, Macdonald, Hernanz \& Jose (1998) found thicker envelope masses
of $q_{\rm He} = 10^{-3} - 10^{-2}$ in a similar analysis, the deeper envelopes resulting from
deeper convection zones because of the use of the OPAL opacities. In our case, our convection
zone is even deeper because of the larger photospheric opacity and consequent change in outer
boundary condition. Thus, we use as our thin helium envelope a mass fraction $q_{\rm He} = 10^{-3.25}$.
As expected, cooling is significantly more rapid for helium atmospheres, because of their lower
opacity. These comparisons are shown in Figure~\ref{LHe}, which
 also shows the $0.61 \rm M_{\odot}$ cooling curve from SDG.
 The model is supposedly based on the standard helium atmosphere models of Wood, but resembles
much more closely our hydrogen atmosphere cooling behaviour. This incongruity exists for all the papers in the
series, suggesting that it is not only the separation energy contribution (which has decreased along the
sequence of papers due to the inclusion of progressively more realistic interior models) but rather a mismatch of atmospheric
models that lies at the root of the discrepant ages. This is important, because the significantly longer
ages found by the European group have been ascribed to their more complete treatment of the crystallisation
process. It appears that much of the delays are actually due to an inaccurate envelope model.

A more direct comparison of envelope models (largely independent of core physics) is to compare
the $L-T_c$ (luminosity-central temperature) relations. This is shown in Figure~\ref{LTc} for the $0.6 M_{\odot}$ models detailed above.
The H \& He sequences initially diverge as the atmospheres become neutral but start to converge
again when the hydrogen opacities become dominated by $\rm H_2$. We see that the Wood hydrogen models
 turn over sooner than the non-grey models. This is due to both the use of a Rosseland mean opacity instead of
a full radiative transfer calculation and also because of the use of Lenzuni, Chernoff \& Salpeter (1991) opacities, which
are not pure hydrogen. To illustrate the difference we also include in Figure~\ref{LTc} a calculation using the same
opacity tables as in our full hydrogen atmosphere calculation, but using only the Rosseland mean opacities. We see
that the turnover is determined primarily by the radiative transfer treatment, while the improved opacity tables become
more important at later times and lower luminosities.
 The $L-T_c$ relation used in the series of
papers culminating in SDG is shown as the long dashed curve and can be found
in Garcia-Berro et al (1996) (a factor 0.6 has been applied to that formula to be appropriate for this
comparison). Here we can see why, although it purports to represent a helium atmosphere
model, the cooling behaviour look more like our hydrogen atmospheres (this fit was based on the
models used by Winget et al 1987, which used opacities from Cox \& Stewart 1965). Recall that $L \propto dT_c/dt$,
so that the deviation shown here results in slower cooling.
Thus cooling models based on this fit 
 are increasingly less accurate for luminosities $\log L/L_{\odot} < -4$.

Why are the helium atmospheres so discrepant? The problem lies with the extreme transparency of neutral
helium. Any approximation which results in an atmosphere containing a small contribution from
another, more opaque material will have dramatic consequences on the atmospheric opacities (Figure~\ref{Phot}
shows the effects of only 1 \% hydrogen by mass). Thus, our helium calculations, which reproduce the photospheric
densities of the BSW atmosphere models, are the first that yield cooling ages appropriate for use with
modern helium atmosphere models.

In the light of these concerns, we must re-evaluate the importance of the release of separation
energies. We calculate a $0.6 M_{\odot}$ white dwarf with the C/O
profile from SDG and both hydrogen and helium atmospheres. In each case we have calculated the
model twice - once with the separation included and concomitant release of energy and once with the
original profile throughout and no energy release. For the case of the hydrogen atmosphere,
 the delay is reduced to $\sim 0.2$~Gyr, and similarly for the helium atmosphere. 
This conforms to what we expect for radiating $\sim 10^{46} \rm ergs$ at $\log L/L_{\odot} \sim -3.5$.
Since, at a given central temperature, our $L-T_c$ relation gives higher L (see Figure~\ref{LTc}),
we expect smaller delays than SDG for a fixed release of energy.
 The reduced influence of separation energy on cooling time is because $\log L/L_{\odot}$ upon
 release is now between -3.5 and -4, as opposed to extending down to -4.5 (e.g. Chabrier 1998).

In summary, we need to explore a range of core compositions to constrain the uncertainties in the models.
To this end we have
calculated our models with hydrogen and helium atmospheres for cores composed of pure Oxygen,
pure Carbon and C/O mixtures given by SDG. Figure~\ref{Cool} shows a sequence of such
models for both helium and hydrogen atmospheres. The much faster cooling of the helium atmospheres
is again evident.

\subsection{Helium Core Models}

Helium core white dwarfs are thought to originate from the truncation of normal stellar
evolution by Roche lobe overflow due to the presence of a binary companion (Kippenhahn, Kohl \& Weigart 1967).
The cooling of such objects has received little attention until the last few years, when the
detections of apparently low mass objects in large surveys (Bergeron, Saffer \& Liebert 1992;
 Bragaglia et al 1990; Bragaglia, Renzini \& Bergeron 1995; Saffer, Livio \& Yungelson 1998) and
low mass companions to millisecond pulsars (e.g. Kulkarni 1986; Lorimer et al 1995; Lundgren et al 1996) 
spurred theoretical efforts (Benvenuto \& Althaus 1998; Hansen \& Phinney 1998a,b).

The cooling of helium core white dwarfs is conceptually similar but contains important differences.
The first is that the heat content stored in the non-degenerate ions ($\propto k T M/Am_p$) is
proportional to the total number of ions for fixed temperature. This means that helium cores
(mass number $A = 4$) contain more heat than carbon or oxygen (A=12 and 16) and thus the helium
core dwarfs are brighter at fixed age. Other more subtle differences result from the lower mass
and hence lower gravities and concomitantly larger non-degenerate layers (and larger radii),
which result in deeper convection zones and cooler atmospheres. Figure~\ref{HeCO} shows the
comparison between our CO and pure helium core sequences. The helium cores are distinctly brighter.

Observed helium white dwarfs usually reside in binaries (however, see Maxted \& Marsh 1998)
and thus  may have potentially observable companions. If the helium dwarf results from the 
original secondary, then it will have an older C/O white dwarf as companion (or a millisecond
pulsar if the original primary was massive enough). Although the helium core dwarfs are brighter,
the presence of a fainter companion can still influence the observed parameters. Consider the
example of a 0.3~$M_{\odot}$ helium core dwarf cooling alongside a companion of $0.6 M_{\odot}$ which
has a cooling age 1~Gyr older (being born from the more massive star). If the pair is unresolved then their
fluxes are simply added. When the helium dwarf is older than 3~Gyr the effective temperatures of
the two stars are similar (and remain so for ages up to $\sim$7~Gyr). Thus, the combined object
will appear to be at the same temperature (since the spectral shape is not sensitive to the gravity)
 although brighter than it should be, yielding an inferred single object mass $\sim 0.21 \rm M_{\odot}$.
This illustrates the problem of interpreting the apparently `overluminous' dwarfs found in proper motion surveys, 
where the presence of a fainter companion can bias the interpretation. Note that this
problem does not arise for millisecond pulsar binaries, such as addressed in Hansen \& Phinney (1998b),
because the dynamical and evolutionary constraints assure the presence of a single body. 

The determination of ages for low mass white dwarfs is further complicated by the possibility
of much larger hydrogen masses on the lowest mass dwarfs (Webbink 1975; Driebe et al 1998) although
this is subject to significant uncertainties in the treatment of wind mass loss and shell flash
burning (Iben \& Tutukov 1986; Driebe et al 1998). One possible constraint on these hydrogen
masses may be obtained by reversing the age arguments in millisecond pulsar binaries to constrain
the ages of the white dwarfs using the pulsar timing ages (Hansen 1999, in preparation).

\section{ Applications}
\label{App}
\subsection{Disk White dwarf Luminosity function I}

The existence of a long suspected (Schmidt 1959) edge to the white dwarf luminosity function 
was demonstrated by Liebert (1979) and Liebert, Dahn \& Monet (1988). Winget et al (1987) derived the first age limits using comparisons
with theoretical models. Thereafter, increasingly sophisticated theoretical models were used by Iben \& Laughlin (1989),
Yuan (1989), Wood (1992), Hernanz et al (1994) and Salaris et al (1997) to further refine the age of the
star forming component of the galactic disk, all using the luminosity function derived by Liebert, Dahn \& Monet
(1988) (hereafter LDM).

The above determinations of an age from the luminosity function are sensitive to several theoretical and observational
uncertainties. The existence of the turnover is well-established, but the theoretical interpretation is sensitive to
the rather uncertain bolometric corrections for the cool white dwarfs at such faint magnitudes. These uncertainties
have been the subject of recent theoretical (BSW) and observational (BRL, LRB) investigations which have
reduced the errors in the LDM sample considerably. However, other proper motion samples of Evans (1992) and
Oswalt et al (1995) have found different turnovers at faint magnitudes, which make the interpretation of
the LDM turnover rather uncertain. Ongoing proper motion surveys in the southern hemisphere (Ruiz \& Takamiya 1995)
also find higher densities of faint white dwarfs than expected from the LDM results.
 Nevertheless, the LDM sample is easily the best studied example and
will form the basis of our analysis below.

There are also several theoretical uncertainties in the age determination, most comprehensively outlined
by Wood (1992).
To determine an age from the luminosity function requires a prescription
for the star formation history, including initial mass function, main sequence-white dwarf mass
function as well as cooling curves (Wood 1992). It also requires an assumption about the relative
populations of hydrogen and helium atmospheres in the white dwarf sample. Most authors assume the
white dwarfs are either exclusively hydrogen atmosphere (Wood 1995) or exclusively helium (Wood 1992; Hernanz et al 1994; SDG).
Below we shall demonstrate how our cooling curves affect results under both assumptions and then
examine how well justified such idealizations are.

We calculate the differential luminosity function (space density per unit luminosity)
$$
\frac{\partial\Phi}{\partial \log L/L_{\odot}} = \int_{M_1}^{M_2} \Psi (t) \xi (M) 
\frac{\partial t_{\rm cool}}{\partial \log L/L_{\odot}} \frac{dM_{\rm wd}}{dM} dM
$$
where $\Psi$ and $\xi$ are the star formation rate and initial mass function respectively.
The white dwarf-main sequence mass relation is $dM_{\rm wd}/dM$ and $\partial t_{\rm cool}/\partial \log L/L_{\odot}$
describes the cooling of the white dwarfs as a function of mass and composition. Our default values
used below include a constant star formation rate, Salpeter mass function and white dwarf-main sequence
mass relation based on that of Wood (1992), $M_{\rm wd} = 0.49 {\rm M_{\odot}} exp(0.096M_{\rm ms})$.
 The cooling time $t_{\rm cool}$ is related to the total stellar
age by $t_{\rm cool}=t-t_{\rm ms}(M_{\rm ms})$ and $t_{\rm ms} = 10 {\rm Gyr} (M/M_{\odot})^{-2.5}$. Our default models
do not include inflation of the stellar scale height with age. To compare with the data, this quantity
must then be integrated over appropriate luminosity bins.

The LDM observational sample was chosen from the Luyten Half Second sample (Luyten 1979). The proper motion
limits are $0.8''{\rm yr^{-1}}<\mu<2.5''{\rm yr^{-1}}$ and for R$<$18. For Galactic disk white dwarfs, the observational
sample is essentially proper motion limited. The turnover in the luminosity function occurs near
$M_{\rm V} \sim 16$, while observations down to $M_{\rm V} \sim 19$ were possible. However, the $V/V_{\rm max}\sim 0.37 <0.5$ for
this sample (LRB), indicating that the sample is not complete.  Based on the detectability of other types
of stars down to $M_{\rm V} \sim 19$, LDM claim that the incompleteness is not severe.
Nevertheless, this makes quantitative estimates somewhat
uncertain. 
 The common proper motion sample of Oswalt et al (1995) (hereafter OSWH) also has $V/V_{\rm max}<0.5$ but contains a 
correction for incompleteness not used by LDM. This results in a somewhat larger effective sample volume
and a more gentle turnover.

Figure~\ref{LFpanels} shows the comparison of our computed luminosity functions (both for pure hydrogen
and pure helium atmospheres) with the two observational data sets. We see that, for hydrogen atmospheres,
the LDM sample yields an age of $8\pm 1$~Gyr and the OSWH sample yields $9.5\pm 1.5$~Gyr. The helium
models yield $5.5\pm 0.5$ and $8\pm 2$~Gyr respectively. These models were calculated with mixed
C/O profiles. Pure Oxygen models reduce the estimates by $\sim$~1~Gyr and pure carbon core models
increase them by $\sim 0.5$~Gyr. The agreement with the conclusions of OSWH is not
surprising given that the non-grey effects which distinguish the current models from
those of Wood only really make a difference below $T_{\rm eff} \sim 5000 K$, corresponding
to $\log L/L_{\odot} < -4.1$. However, for questions regarding dwarfs of greater age
than that of the peak, the difference will become appreciable.
 The faster cooling of helium atmospheres results in a much
broader peak in the luminosity function and a fit to much younger disk ages, than previously
derived. Thus, previous ages based on helium atmosphere models are
 considerably overestimated.

We have seen that the choice of atmosphere composition makes a difference $\sim 2$~Gyr in the inferred
ages. 
 We know that both types are present in the number counts, so it makes sense to examine
hybrid luminosity functions. As a prelude to that, we now examine the properties of
the individual dwarfs that define the peak and turnover of the LDM luminosity function.

\subsection{Individual Old White Dwarfs}

The application of detailed atmospheric models has led to the determination of
individual gravities and effective temperatures for many of the cooler dwarfs
(BRL), including those in the LDM sample (LRB). The determination of masses
and compositions  allows us to investigate the ages of individual stars.

The most interesting stars are obviously those that define the turnover
of the LDM  luminosity function. We will consider the bolometric luminosity function 
given by LRB (the different bolometric corrections for hydrogen and helium atmospheres
means that different stars may inhabit the faintest luminosity bin for $\rm M_V$ and $\rm M_{bol}$ luminosity
functions).
 The faintest luminosity bin defined by LDM
contained three stars, 1300+263, 2251-070 and 2002-110.
All three are classified as helium atmosphere dwarfs, with masses determined
by BRL and LRB to be $M=0.72\pm 0.11 {\rm M_{\odot}}$, $M = 0.82 \pm 0.03 {\rm M_{\odot}}$ 
and $M = 0.78 \pm 0.01 {\rm M_{\odot}}$ respectively.
The effective temperatures  are  all between $4500-5000$~K which yield ages of $4.8\pm 0.2$,
$4.7\pm 0.2$ and $4.5 \pm 0.1$~Gyr respectively. These are considerably less than those quoted by BRL and LRB
because they used the helium atmosphere models of Wood, whose outer boundary
conditions are not consistent with
the atmospheric determinations at these low temperatures. 

The next coolest bin  contains 8 stars, 5 helium and 3 hydrogen atmospheres. However, 2 of the hydrogen
atmosphere dwarfs have low inferred gravities, suggesting that they may be helium core white dwarfs
with masses $\sim 0.3-0.4 \rm M_{\odot}$. Another possible interpretation is that they are unresolved binaries,
which thus appear overluminous (and yield a lower gravity in the BRL and LRB analyses). Interpreting
these two objects as either individual helium dwarfs or as an unresolved binary containing
  two identical more massive dwarfs (of same effective temperature as inferred)  yields
ages $\sim 10$~Gyr in both cases 
The only `normal'  hydrogen atmosphere dwarf in this bin is  1108+207, with
a mass $0.63\pm 0.07 \rm M_{\odot}$ and inferred age $6.3 \pm 0.8$~Gyr. All the helium atmosphere dwarfs in 
this bin have ages between $4-5$~Gyr.
 Figure~\ref{DA} and Figure~\ref{DB} show all the dwarfs in the LDM sample, with parameters
taken from LRB, along with appropriate cooling curves and age curves. Table~\ref{LDMTAB} shows
the individual ages for the faintest objects.

Several features should be noted about Figures~\ref{DA} and \ref{DB}. The early crystallisation
and subsequent rapid cooling of the more massive models causes the isochrones to `bend over' at
the top of the diagram, as noted by several authors before.
 Note also the discontinuous nature of the isochrones between C/O and
helium cores, which is caused by the greater heat capacity of a core composed of smaller ions,
thereby making the helium core dwarfs brighter than C/O dwarfs of comparable mass. 

Figure~\ref{DA} shows that the hydrogen atmosphere dwarfs also show a concentration of stars with ages $\sim$4~Gyr, with only 1108+207 of the
`normal' dwarfs having an age $>$5~Gyr. These stars were not mentioned above because they are brighter than the helium dwarfs i.e. 
the fainter magnitude bins are dominated by helium atmosphere dwarfs. Indeed, of the 11 stars in the faintest two bins, 8 were helium
atmosphere dwarfs, despite the fact that the LDM sample contains approximately equal numbers of hydrogen and helium atmospheres (22 hydrogen
and 20 helium).
 The helium atmosphere dwarfs in Figure~\ref{DB} also show
a concentration at similar ages, although at lower effective temperatures, because of their more
rapid cooling. Thus, the `non-DA gap' identified by BRL is simply a result of the different
cooling rates, rather than due to any chemical evolution. These figures also suggest that
the ages inferred from our helium model luminosity function are more accurate, primarily because
the fainter magnitude bins contain more helium atmosphere dwarfs than hydrogen dwarfs.
 The
 ages of the apparently overluminous hydrogen atmosphere dwarfs are discrepant with this picture
but have to be regarded with caution because they may be unresolved binaries and 
 the combination of two different atmospheres is
a non-linear process and makes the inversion to determine the parameters rather difficult.

With individual mass determinations we may now simply use the oldest white dwarf as a constraint
on the galactic disk age.
Neglecting the possibly overluminous dwarfs, the oldest dwarf in the LDM sample is 1108+207, with an age
$6.3 \pm 0.8$~Gyr.
 For individual dwarf constraints we need not bother with completeness concerns and
we  can turn to the larger sample of BRL,
 where the oldest
star is the hydrogen atmosphere dwarf 1310-472, which has mass 0.63$\pm 0.03 \rm M_{\odot}$ and
 age 8.4$\pm$0.3~Gyr
(white dwarf cooling age only).  If one considers the apparent low mass dwarfs at face value as helium 
core dwarfs, the oldest of these
is 0747+073A, with a mass $\sim 0.4 \rm M_{\odot}$ and an age $10.8 \pm 0.4$~Gyr.

\subsection{Disk White Dwarf Luminosity Function II}

The above is a very simple constraint and there is much more information to be gained from applying a detailed
analysis to the full luminosity
function. However, we have demonstrated that such an exercise really must be done for both hydrogen and helium atmospheres
separately. The fact that both DA and DB samples have concentrations of white dwarfs at similar ages $\sim 4.5$~Gyr, although
at different temperatures, supports the claim that these two populations of white dwarfs really are distinct and
remain that way, cooling at different rates. This fact must also be taken into account when attempting to analyse the
chemical evolution of white dwarfs.
Thus, the theoretical luminosity function must take account of this differential cooling.
 Figure~\ref{LFDAB} shows the comparison of the data with a luminosity function in
which it is assumed that 50 \% of all white dwarfs are born with hydrogen and 50 \% with helium atmospheres which then cool
according to their respective cooling tracks. While this is obviously in conflict with observations of more luminous
white dwarfs, it is consistent with the observations for white dwarfs with $T_{\rm eff} < 10^4$~K, i.e. over the entire
LDM sample (and most of the cooling age is spent in this range).
 In this case, comparison with the observations 
provides an estimated disk age of $6.5\pm 1$ Gyrs for the LDM sample. The more gradual
turnover of the OSWH sample is consistent with a wide range of ages from 6-11~Gyr. The inclusion of
both kinds of white dwarfs
allows for a peak in the luminosity function at $\log L/L_{\odot} \sim -4$ (as observed) while maintaining
a more gradual turnover than would occur with pure hydrogen models. This provides an acceptable fits over
a wide range of ages.

Another test of the white dwarf models and stellar evolution history used to construct the luminosity function
 is to compare the observed fraction of helium to hydrogen white dwarfs in different magnitude bins. Using the simple
50/50 split with constant star formation rate assumed in Figure~\ref{LFDAB} we have calculated the `DB fraction' in 4 luminosity
bins. This is shown in Figure~\ref{DAB}.
 The theoretical curves of different ages are compared to the weighted fractions calculated
for the LDM sample using the $M_{bol}$ and $1/v_{max}$ weights of LRB. The error bars are obtained by removing the single
largest contributor to either helium and hydrogen atmosphere fractions in each magnitude bin. We see that the general trend
of rising helium fraction at low luminosities is consistent with the 6 Gyr curve although the brightest point appears
discrepant with all theoretical curves. The equivalent numbers from the OSWH sample are not shown for two
reasons. The first is that the atmospheric compositions have not been studied in the same detail (making assignment
to hydrogen and helium bins less certain) and the second is
that the large completeness corrections used by Oswalt et al (1995) result in each luminosity bin being dominated by a single
object. Given the uncertainty in the completeness of the LDM sample, the significance of the agreement is not clear.
However, we have seen that it is essential to take account of both hydrogen and helium cooling rates separately,
and such a comparison offers the possibility of constraining the relative populations of hydrogen and helium
in future analyses.

To conclude, the comparison of the models presented here
with the most well-studied white dwarf proper motion sample (LDM) suggests an age
of only $\sim 6.5\pm 1$~Gyrs for the stellar populations in the solar neighbourhood, somewhat less than 
other recent determinations. This difference is primarily due to our updated Helium atmosphere opacities, because the
faintest stars in the LDM sample are of this type.
 However, the completeness
of the LDM sample at faint magnitudes has been questioned (Ruiz \& Takamiya 1995; Oswalt et al 1995) and so
this should really only be considered a lower limit.
Comparison with the
  OSWH sample of Oswalt et al (1995) suggests a much larger age range.
 However, the
white dwarfs in these systems have not been studied to the same detail as those in the LDM sample and, in some cases,
 uncertainties exist about their exact luminosities and chemical compositions (see LRB). Nevertheless, this age
is consistent with the inferred ages of individual helium core objects in the LDM sample as well as the oldest
hydrogen atmosphere C/O core dwarfs studied by BRL.
 Clearly a larger proper
motion sample is needed to affirm these conclusions.

\subsection{Pulsar-White Dwarf Binaries}

A different sample of white dwarfs in binaries are the low-mass helium-core companions to millisecond pulsars.
The well-determined parameters of these binaries preclude the possibility of close, blended identical dwarfs, i.e. these
really are helium core white dwarfs. Hansen \& Phinney (1998b) have determined cooling ages for several of these stars.
However, the greater distances to these systems mean that accurate distance measures are often unavailable.
 In such cases, distance information is based on the dispersion measure of the companion pulsar.
Despite these uncertainties, meaningful constraints on the
cooling ages may be obtained in several cases, especially in those cases where pulsar timing can
indeed provide meaningful distance
limits. 
The oldest white dwarf in this sample is the companion to the pulsar
PSR J1713+0747, whose cooling age is $>5.2$~Gyr\footnote{The value quoted in Hansen \& Phinney is 6.3 Gyr and was obtained using
hydrogen atmosphere models. Recent more sophisticated models (Hansen 1998a), including helium atmospheres, allow a slightly lower cooling age}.
 Most of the others have lower age limits in the range 3-5~Gyr, although the upper limits often extend to $>10$~Gyr.
 Thus the age limits from these binaries 
are consistent with the low Galactic disk ages derived for the LDM sample. However, these age estimates were based
on optical observations only. As emphasized by BRL, infra-red observations can offer much extra information
and may help to constrain these white dwarfs further.

\subsection{Halo White Dwarfs}

White dwarfs have been considered as baryonic dark matter candidates for many
years (Larson 1986; Silk 1991; Carr 1994) but have received special attention
recently resulting from the observation of  microlensing towards the
Large Magellanic Cloud (Paczynski 1986; Alcock et al 1997). This has resulted in several
papers attempting to constrain the white dwarf halo scenario using various observations of
the presumed local contribution (Adams \& Laughlin 1996; Chabrier, Segretain
\& Mera 1996; Graff, Laughlin \& Freese 1998; Isern et al 1998). Unfortunately, none of these
papers uses white dwarf models which remain accurate to the ages required to produce
a proper constraint. In particular, the last three used the results of Hernanz et al,
which we have seen are very inaccurate for halo ages.

The two observational data sets commonly used are the halo contribution to the
LDM white dwarf luminosity function and the Hubble Deep Field (Williams et al 1996) point source number
counts (Flynn, Gould \& Bahcall 1996, Elson, Santiago \& Gilmore 1996; Mendez et al 1996).
As we have noted above, the completeness of the white dwarf luminosity function at faint
magnitudes is somewhat uncertain, making inferences based
on non-detections particularly troubling. Nevertheless, it is illustrative to examine the
constraints.

As noted by Graff et al (1998), the larger relative velocities of the halo sample make the
upper cutoff in proper motion (2.5$''$/yr) quite important. As the white dwarfs fade, the
magnitude limited distance starts to approach the minimum distance set by this upper proper
motion cutoff, leaving little volume available for detection. To examine the constraints
imposed by the LDM luminosity function, let us first assume it is complete to the stated
limit of R = 18. We shall adopt the velocity distribution used by Alcock et al (1997) in
deriving the characteristic MACHO mass, namely a 3-D Maxwellian with isotropic velocity
dispersion $\sigma = 150 \rm km/s$\footnote{ 
Graff et al (1998) state that they
use a dispersion $\sigma$=270 km/s. However, this is their 3-D rms velocity, so that
their model is actually the same as that used by by the MACHO group (D.Graff, personal communication)}.
 Using a circular velocity of
220~km/s, we determine the effective volume probed by the LDM survey for white dwarfs of different
age and atmospheric composition.
We shall
constrain the halo fraction in white dwarfs by requiring that a halo fraction $f_{\rm halo}$ predict at least
3 white dwarfs in a given magnitude bin (to assure detection in the face of Poisson fluctuations).
 We use a local dark matter density $\rho_{\odot} = \rm
0.0078 M_{\odot} pc^{-3}$. We will also assume 0.7~$\rm M_{\odot}$ objects as a conservative estimate,
since Weidemann (1987) suggests that the lower metallicity population of the LMC may give rise
to higher mass white dwarfs and we should expect a similar trend here\footnote{However, this is
far from conclusive. Richer et al (1997) find the white dwarf sequence in the old globular cluster
M4 to be consistent with 0.51$\rm M_{\odot}$}.
 Thus, the halo mass fraction is constrained to be
$$
f_{\rm halo} < \frac{0.008}{(D_{\rm eff}/20{\rm pc})^3}
$$
where $D_{\rm eff}$ is the radius of the effective spherical volume probed by the LDM survey at
the given magnitude. Figure~\ref{fhalo} shows the value of $f_{\rm halo}$ for hydrogen and helium
atmosphere dwarfs of various ages. We see that the latter are consistent with the LDM non-detections
for all ages $>$7.5~Gyr. Hydrogen white dwarfs are more strongly constrained, because of their
slower cooling. If we assume a 50/50 split between hydrogen and helium atmospheres as before, the LDM
constraints are consistent with $f_{\rm halo}$=0.3 for pure Oxygen core models
 and white dwarf ages $>14$~Gyr. If the white dwarfs are all helium atmosphere dwarfs, then ages of
only 7.5~Gyr are required.

The above assumes the LDM sample is complete, despite our suspicions to the contrary. The constraints
are obviously lessened if incompleteness is accounted for. As 
 a simple estimate of the effects of incompleteness, 
we note that both Oswalt et al (1995) and Ruiz \& Takamiya (1995) find the faintest
white dwarfs to be $\sim 5$ times more numerous than suggested by LDM, so let us adopt a uniform completeness of 20 \%.
Although this was chosen for simplicity, there is some support for such a uniform correction. The fact that
LDM did detect stars of other kinds down to $M_{\rm V}=19$ suggests that any incompleteness in the white dwarf
luminosity function is not a strong function of magnitude at the faint end.
Thus, with this incompleteness, the required value of $f_{\rm halo}$ in Figure~\ref{fhalo} should be reduced by a factor 5.
 This allows a somewhat weaker constraint on 
hydrogen white dwarfs $\sim$11.5~Gyr.
There are several
uncertain assumptions underlying these numbers, but they serve to illustrate that, given the uncertainties
in the completeness of the LDM sample at faint magnitudes, it is quite possible to make a white
dwarf halo model consistent with the data. Given the blueward shift of the hydrogen dwarfs at late
times, one must also worry about colour selection in these proper motion samples. This is probably
not a problem for studies of the disk luminosity function, because the faintest magnitude bins are
dominated by helium atmosphere dwarfs (which remain red). 

An alternative model for the microlensing results invokes a dark thick disk or spheroid (Gates et al 1998). 
The greater concentration towards the disk allows one to obtain the required optical depth to microlensing
with a smaller fraction of the dynamical halo mass. The estimated mass of the lenses is still
 $\sim 0.2-0.4 M_{\rm \odot}$
which again suggests some kind of stellar relic. To place limits on such a population, we shall consider
a population of white dwarfs which rotate with the circular velocity but have velocity dispersions of 50 and
135 km/s (to match the Gates et al scale heights of 1.5 and 2.5 kpc respectively). These limits are
also shown in Figure~\ref{fhalo}. Note that $f_{halo}$ is actually the fraction of the local dark matter density
in old white dwarfs, so that the constraints can also be applied to this case. We see in Figure~\ref{fhalo} that
the $50$~km/s model is ruled out for ages $<$15~Gyr.

The constraints from the Hubble Deep Field (Williams et al 1996) are weaker but more robust
(Hansen 1998b).
The HDF is a very deep, but extremely narrow spatial sample. Since we have no proper motions
for the detected objects,
we have to rely on number counts and colour selection to constrain the white dwarfs. 
Figure~\ref{HDF1} shows the $V-I$ versus $V$ plot for all the point sources identified by
Flynn et al (1996), Elson et al (1996) and Mendez et al (1996). Also plotted are the expected positions of
 $0.6 M_{\odot}$ white dwarfs of hydrogen and helium atmospheric compositions.
 We see that the HDF does not provide any meaningful constraints
on helium atmosphere dwarfs (because of their rapid cooling). Hydrogen atmosphere dwarfs
should be observable within $\sim 2 {\rm kpc}$, and their colours in fact correspond well with
the main group of faint blue point sources detected by Elson et al (1996). This population was
originally ruled out because of the assumption that all old white dwarfs become redder
with age, whereas we have shown above that hydrogen white dwarfs eventually turn towards
the blue again because of the molecular hydrogen in their atmospheres.

Having determined that white dwarfs could be present, we also need to determine whether
there are enough point sources. Helium dwarfs cool too rapidly to be of interest down to
the magnitude limit, so we consider only hydrogen atmosphere dwarfs.
 The HDF covers a field of view of $\Omega =4.4$ square arcminutes.
Thus, the volume probed out to a given distance $d$ determined by a magnitude limit of
$m_V$ is
\be V = \frac{\Omega}{3} 10^{0.6 (m_V-M_V)+3} pc^3, \ee
where $M_V$ is the absolute magnitude of the population of objects.
Given a local density $\rho$ of dark matter, and a characteristic dwarf mass $m_{\rm wd}$, we 
may predict a number
\be N \sim \frac{f \rho}{m_{wd}} \frac{\Omega}{3} 10^{0.6 (m_V-M_V)+3} \ee
of objects above a given magnitude limit (where $f$ is the fraction of the dark mass
stored in such objects). Assuming $\rho \sim 0.0078 \rm M_{\odot} pc^{-3}$ as above, 
completeness limit $m_{\rm V} \sim 28$, and a hydrogen white dwarf population of
age $\sim 14$~Gyr ($M_{\rm V} \sim 17.7$), we have $N \sim 2.1 f$. Given that the microlensing
results suggest $f \sim 0.5$ and that we are considering only the hydrogen atmospheres (estimated
to be $\sim 0.5$ for disk dwarfs), this implies $N \sim 0.53 $. Since several blue point sources
were detected,
  the HDF doesn't place much of
a constraint on very old hydrogen atmosphere dwarfs either. A similar conclusion was reached
by Graff et al (1998), who found that the constraints from the LDM sample were more restrictive. However,
as we saw above, concerns exist about the completeness of the LDM sample. The HDF results suggest
that proper motions are essential to detect old hydrogen atmosphere dwarfs, since there will be an
abundance of extragalactic objects of
similar magnitude and colour in deep images (Elson et al 1996; Hansen 1998b).

\subsection{Globular Clusters}

The existence of a 12~Gyr-old white dwarf dark halo is still somewhat
uncertain. However, the mean age of the Galactic globular cluster system is estimated to be 
$11.5 \pm 1.3$~Gyr (Chaboyer et al 1998), so that the white dwarf sequences in these objects offer the potential
to test the cooling curves derived above and verify the constraints they place on the detection
of dark halo dwarfs as well as offering another method for globular cluster age determination.
The large deviations from black body colours shown by old hydrogen atmosphere
white dwarfs suggests that sufficiently deep observations
of the white dwarf sequence in globular clusters should make it possible to observe a splitting
in the white dwarf cooling sequences near $M_{\rm V} \sim 17$ into hydrogen and helium parts.
 Interestingly, the observations of
M4 (Richer et al 1997) reach almost this far (to white dwarf ages $\sim 9 \rm Gyr$). With deeper
observations, it should be possible to examine this splitting, which may provide a more tangible
estimate of white dwarf age than a simple cutoff, whose veracity is a function of completeness
(see Figure~\ref{M4}).
Furthermore, given the very small amount of gas found in globular clusters, the accretion history
of globular cluster dwarfs is likely to be quite similar to that of halo dwarfs. If accretion of
interstellar material has any effect on the relative abundances of hydrogen and helium in
white dwarf atmospheres,
 the determination 
of relative proportions of hydrogen and helium atmospheres amongst the globular cluster
dwarfs will be useful in comparing the disk population to the putative halo population.

However, the exact nature of the hydrogen cooling tracks will be sensitive to the
main sequence-white dwarf mass relation. For a population of coeval stars, more massive
white dwarfs are born first and will thus have longer cooling ages. Thus, the white dwarf
mass will vary along the cooling
sequence in a globular cluster. For the purposes of current comparisons, this is not a
large source of error as the upper parts of the cooling sequence are essentially the
same mass. However, this mass variation must be taken into account when it becomes 
possible to examine the splitting of the cooling sequences. Figure~\ref{GC} shows the
cooling sequence for populations that are 11 and 13 Gyr old, using the initial-final
mass relation of Wood (1992). Also shown is the cooling sequence using only the mass of 0.53$M_{\odot}$
determined by Renzini et al (1996) as appropriate for the upper part of the sequence of
the cluster NGC6752. The significant deviations in the nature of the white dwarf turnoff are apparent.

\section{Summary}

This paper describes the construction of a new set of white dwarf cooling models appropriate for
the study of very old white dwarfs. The principal advance is the incorporation of new opacities
and  an accurate
outer boundary condition based on proper radiative transfer calculations in the white dwarf
atmosphere. These models allow us to study in detail the differences in cooling between dwarfs
with hydrogen and helium atmospheres and to extend the cooling sequences to ages $\sim 15$~Gyr,
suitable for studying the oldest white dwarfs in our Galaxy. 
Our models also incorporate a variety of core compositions,
which allow us to compare cooling ages for low mass and high mass white dwarfs. 

We have used the new models to study several important questions concerning white dwarf
evolution and its consequences. The new cooling curves reduce the possible importance
of the separation energy (Garcia-Berro et al 1988; Segretain et al 1994; SDG) because the
energy is released at higher luminosities, hence the delay introduced by a fixed energy release is smaller.
 We have
also shown that the `non-DA' gap between 5000-6000~K identified by Bergeron, Ruiz \& Leggett (1997) may simply
be a consequence of the different cooling rates of hydrogen and helium atmospheres, because the hydrogen
atmosphere dwarfs above the gap are of the same age as the helium atmosphere dwarfs below the gap.

We can place constraints on the age of the Galactic disk using the luminosity function of local
proper motion stars (Liebert, Dahn \& Monet 1988; Leggett, Ruiz \& Bergeron 1998) and
common proper motion binaries (Oswalt et al 1995) using the new models. The LDM sample yields
a very small age $\sim 6.5 \pm 1$~Gyr if one assumes that the sample is complete. This is
also consistent with the age determinations for individual stars in the sample, except for
two `overluminous' hydrogen atmosphere dwarfs which may be low mass helium-core dwarfs.
The latter yield ages $\sim 10-11$~Gyr, which is larger than the LDM limits, but
consistent with the wider allowed age range we derive from OSWH luminosity function and individual
ages for some
dwarfs not in the LDM sample.
Our models are consistent with ages considerably less than some derived recently in
the literature, primarily because of our updated opacities and the recognition that Helium
atmosphere dwarfs cool much faster than Hydrogen atmosphere dwarfs.
 However, the determination of disk ages from the white dwarf luminosity
function will not be conclusive until several outstanding issues regarding the completeness
of the luminosity function at the faint end have been addressed. This may also be linked to
the kinematic evolution of the white dwarf population (Garcia-Berro et al 1998).

Despite uncomfortable chemical evolution constraints (Gibson \& Mould 1997; Fields, Mathews \&
Schramm 1997), white dwarfs remain a favoured candidate for halo dark matter. We have used
our new models to demonstrate that previous attempts to constrain such populations directly
(Adams \& Laughlin 1996; Chabrier, Segretain \& Mera 1996; Graff, Laughlin \& Freese 1998; Isern et al 1998)
were too severe in their constraints because of inaccuracies in the cooling models. We have
presented conservative estimates of the constraints on such models and find that it is possible
to make consistent white dwarf halo models with ages no larger than those of the globular
clusters (based on direct detection criteria; chemical evolution constraints persist). We have
also presented predictions for the colours and luminosities of the putative population. 
For white dwarfs older than $\sim 9 Gyr$, a full radiative transfer treatment of the outer
boundary condition is very important.

Deep proper motion surveys for old white dwarfs can add much to our
current understanding of Galactic structure and formation. However, to obtain the most
information, it will be important to determine the white dwarf atmospheric composition
through spectroscopy and/or multi-wavelength photometry and to calculate cooling
models that are consistent with the derived parameters. We also note that the photometric
peculiarities of old hydrogen atmosphere dwarfs must be accounted for in colour selection of
observational samples.

In conclusion, I would like to thank Matt Wood, Harvey Richer, Sterl Phinney, Aleksandra Borysow, 
 Gilles Chabrier, Margarita Hernanz, Enrique Garcia-Berro and the referee Don Winget 
for comments, exchanges of results and/or encouragement at various stages of this work. 
This work is supported by NSERC of Canada.

\clearpage


\figcaption[match.ps]{The open symbols represent the Rosseland mean opacity from the OPAL calculations.
The filled symbols are the equivalent from our calculations while the stars are the Planck mean from
our calculation. The vertical dotted line indicates the region of overlap between our calculation 
and the OPAL calculation. The deviation between Rosseland and Planck means below 5000~K is a diagnostic
of important non-grey effects in the atmosphere. \label{RossPlanck}}

\figcaption[photop.ps]{This figure shows the main opacity contributions at the photosphere
for both a pure hydrogen ($\rho \sim 10^{-2} g/cm^3$) and pure helium ($\rho \sim 100 g/cm^3$)
atmosphere at $T_{\rm eff} = 4000$~K. The middle panel shows the Planck function at this 
temperature, which indicates the distribution of flux with wavelength before modification
by the radiative transfer. The powerful influence of the molecular opacities for $\lambda > 1 \mu m$
is evident in the top panel.
 \label{Photop}}

\figcaption[pt.ps]{ The solid and dotted lines indicate the positions of the photosphere
for a pure hydrogen and pure helium atmosphere respectively. The dashed lines indicate the
same for mixed H/He atmospheres with the indicated mass fractions of hydrogen. The closed and
open circles indicate the corresponding quantities from BSW. The label
PPT indicates the location of the Plasma Phase Transition of Saumon \& Chabrier (1992), which corresponds
to the pressure ionisation of hydrogen. Thus, a correct treatment of pressure ionisation is not
 critical to the calculation of hydrogen atmospheres, although it is very important for cool
helium atmospheres.  \label{Phot}}

\figcaption[bands4000.ps]{ The heavy solid line is the emergent spectrum for a pure hydrogen atmosphere and
the thin solid line is for a pure helium atmosphere. The dotted line is the Planck function (all of these
spectra are for 4000 K, so this Planck function corresponds to the one in Figure~\protect{\ref{Photop}}).
The dashed line is the emergent spectrum for the mixed H/He atmosphere which 10 \% hydrogen by mass. \label{Bands}}


\figcaption[massH.ps]{The curves show the  hydrogen mass fraction $q_H$ as a function
of time for models of several masses, all starting with $q_H=10^{-4}$. The more massive stars burn
more hydrogen because of larger temperatures and pressures at the base of the hydrogen layer.\label{masses}}

\figcaption[LH.ps]{ Here we compare hydrogen atmosphere cooling models. The solid lines are our models for Carbon and Oxygen cores respectively (Carbon being the
brighter at late times). The dotted lines are the equivalent models from Wood (1995). The dashed line is our model
using the C/O mixture outlined in the text.\label{LH}}

\figcaption[LHe.ps]{Here we compare helium atmosphere cooling models. The solid lines are our models for Carbon and Oxygen interiors respectively, while the dashed line
is for the C/O mixture.  The filled circles are the results from Salaris et al (1997) for the same C/O mixture while
the dotted line is the result of Wood for pure Carbon cores.\label{LHe}}

\figcaption[LTc.ps]{ The solid lines are our models for pure hydrogen (top) and
pure helium (bottom) atmospheres. The dotted lines are the equivalent models by Wood.
The points represent a cooling model calculated with a frequency averaged version of our
opacity tables.
The long dashed line is the fit from Garcia-Berro et al (1996) adjusted to be appropriate
for this comparison. The marked differences in these curves explain most of the differences in cooling ages.\label{LTc}}


\figcaption[Cool.ps]{The top panel shows the cooling for C/O core models with hydrogen atmospheres. The
bottom panel shows the same core models with helium atmospheres instead. The more massive models crystallise
first and therefore cool faster at later times. \label{Cool}}

\figcaption[HeCO.ps]{The solid lines show C/O core models from 0.5 $M_{\odot}$ to 0.9 $M_{\odot}$.
The dashed lines show helium core models from 0.3-0.45 $M_{\odot}$. All models have hydrogen
atmospheres. \label{HeCO}}

\figcaption[LFpanels.ps]{ The left hand panels show the comparison of the Liebert et al data
with our hydrogen (top) and helium (bottom) atmosphere luminosity functions. The open circles
are the data of Fleming et al.
The right hand
side shows the same comparison but with the Common Proper Motion data of Oswalt et al. The
curves for the hydrogen models are for disk ages of 6,8,10 and 12 Gyr. The helium curves include
a 4~Gyr curve as well.
 \label{LFpanels}}

\figcaption[DA.ps]{The hydrogen atmosphere dwarfs shown here are those in the LDM sample. The dotted lines
are cooling curves for C/O core models and the short dashed lines are for helium core models. The solid
lines are isochrones while the two long dashed curves indicate the region over which core crystallisation
takes place. The concentration of stars in the region near 4~Gyr is apparent. \label{DA}}

\figcaption[DB.ps]{The helium atmosphere dwarfs shown here are from the LDM sample. Once again, the dotted
lines indicate cooling curves for C/O models and the solid lines are isochrones. The dashed lines indicate
the region over which core crystallisation takes place. The non-DA gap alluded to by BRL is evident here
between 6000 and 5000 K. \label{DB}}


\figcaption[LFDAB.ps]{ The luminosity functions shown here contains coeval populations of $50\%$ of hydrogen atmosphere
dwarfs and $50\%$ helium atmosphere dwarfs. The upper panel shows the comparison with the LDM sample and
the lower panel the comparison with the data of OSWH. All curves are shown with the same star formation rate, but this can be
varied within the bounds of the data to determine the full range of consistent ages.
 \label{LFDAB}}

\figcaption[DAB.ps]{The 4 curves are for disk ages of 4,6,8 and 10 Gyr. The four plotted points are the fraction inferred
from the LDM sample with the number of stars in each bin shown. \label{DAB}}

\figcaption[fhalo.ps]{ The filled circles represent the hydrogen atmosphere constraints and the open
circles the helium atmosphere constraints. All models are for $0.7 M_{\odot}$ pure Oxygen cores. The 
solid line is for the MACHO models and the dashed lines for the Gates et al (1998) models (the upper curve
has the higher velocity dispersion in each case). The upper shaded region is the halo fraction
claimed by the Alcock et al (1997) assuming the LDM sample is complete and the halo is formed entirely
from either helium or hydrogen dwarfs. The bottom shaded region assumes the LDM sample is 20\% complete
and the halo is 50 \% hydrogen atmospheres and 50 \% helium. The estimated mean age of the globular
cluster system from Chaboyer et al (1998) is also shown.  \label{fhalo}}

\figcaption[HDF2.ps]{The solid curves are hydrogen atmosphere models at distances of 1~kpc and 2~kpc.
The dashed lines are the equivalent helium atmosphere models. The solid and open points are the
point sources from Elson et al (1996) and Mendez et al (1996). The dotted line encloses the `halo region'
of Flynn et al (1996). The curves are labelled with ages in Gyr.
\label{HDF1}}


\figcaption[M4.ps]{The HR diagram for the globular cluster M4 (Richer et al 1997) shows
an extensive white dwarf cooling track. Shown on here are two tracks for a 0.6 $M_{\odot}$ dwarf
with C/O composition and hydrogen (solid) and helium (dotted) atmospheres. \label{M4}}

\figcaption[GC.ps]{The solid lines are coeval white dwarf cooling sequences of ages 11 and 13 Gyr respectively. The dashed line is the cooling curve for $0.53M_{\odot}$ white dwarfs. On the right is
shown the white dwarf masses as a function of $M_V$ for the 13 Gyr sequence.\label{GC}}

\begin{deluxetable}{llcclcllc}
\tablewidth{0pc}
\tablecaption{Ages for old LDM dwarfs \label{LDMtab}}
\tablehead{
\colhead{Name}   &
 \colhead{$\pi$} & \colhead{V} & \colhead{V-I} & \colhead{H/He} &
 \colhead{M($M_{\odot}$)} & \colhead{$T_{\rm eff}$} &
\colhead{Age (Gyr)} & \colhead{$\Delta t$ (Gyr)} 
}
\startdata
1300+263 & 28 $\pm$3 & 18.77 & 1.28 & He & 0.72$\pm$ 0.11 & 4539$\pm$50 & 4.8$\pm$0.2 & $-$4.7
 \nl
2251-070 & 124$\pm$4  & 15.71  & 1.15 & He & 0.82$\pm$0.03 & 4590$\pm$70  & 4.7$\pm$0.1 & $-$5.0
 \nl
2002-110 & 58$\pm$1 & 16.95 & 1.09 & He & 0.78$\pm$0.01 & 4813$\pm$54 & 4.5$\pm$0.1 & $-$4.5
 \nl
\hline
1247+551 & 40$\pm$1 & 17.79  & 1.45 & H & 0.33$\pm$0.01 & 4000$\pm$70  & 9.8$\pm$0.3$^{*}$ & +5.7
\nl
1108+207 & 38$\pm$3 & 17.70 & 1.07 & H & 0.63$\pm$0.07 & 4640$\pm$160 & 6.3$\pm$1.0 & $-$1.0
\nl
0747+073A & 55$\pm$1 & 16.96 & 1.26 & H & 0.39$\pm$0.01 & 4166$\pm$81 & 10.8$\pm$0.4$^{*}$ & +6.1
 \nl
2316-065 & 32$\pm$4 & 18.15 & 1.16 & He & 0.69$\pm$0.11 & 4747$\pm$53 & 4.7$\pm$0.1 & $-$3.9
\nl
0552-041 & 155$\pm$2 & 14.47 & 0.98 & He & 0.78$\pm$0.01 & 5080$\pm$60 & 4.3$\pm$0.1 & $-$3.9
\nl
0747+073B & 55$\pm$1 & 16.63 & 1.08 & He & 0.60$\pm$0.01 & 4871$\pm$54 & 4.1$\pm$0.1 & $-$2.7
\nl
2054-050 & 65$\pm$5 & 16.69 & 1.32 & He & 0.63$\pm$0.08 & 4630$\pm$50 & 4.5$\pm$0.1 & $-$3.8
\nl
1444-175 & 69$\pm$4 & 16.44 & 1.01 & He & 0.83$\pm$0.05 & 4990$\pm$60 & 4.3$\pm$0.1 & $-$4.2
\nl

\tablecomments{These values assume mass fraction q$_{H}$=$10^{-4}$ for hydrogen atmospheres and 
q$_{He}$=$10^{-3.25}$ for helium atmospheres. The last column indicates the age difference we infer
here with respect to that given in table 2 of Leggett, Ruiz \& Bergeron (1998). Those ages marked
with an asterisk assume that the observations are uncontaminated by companion light.\label{LDMTAB}}
\enddata
\end{deluxetable}

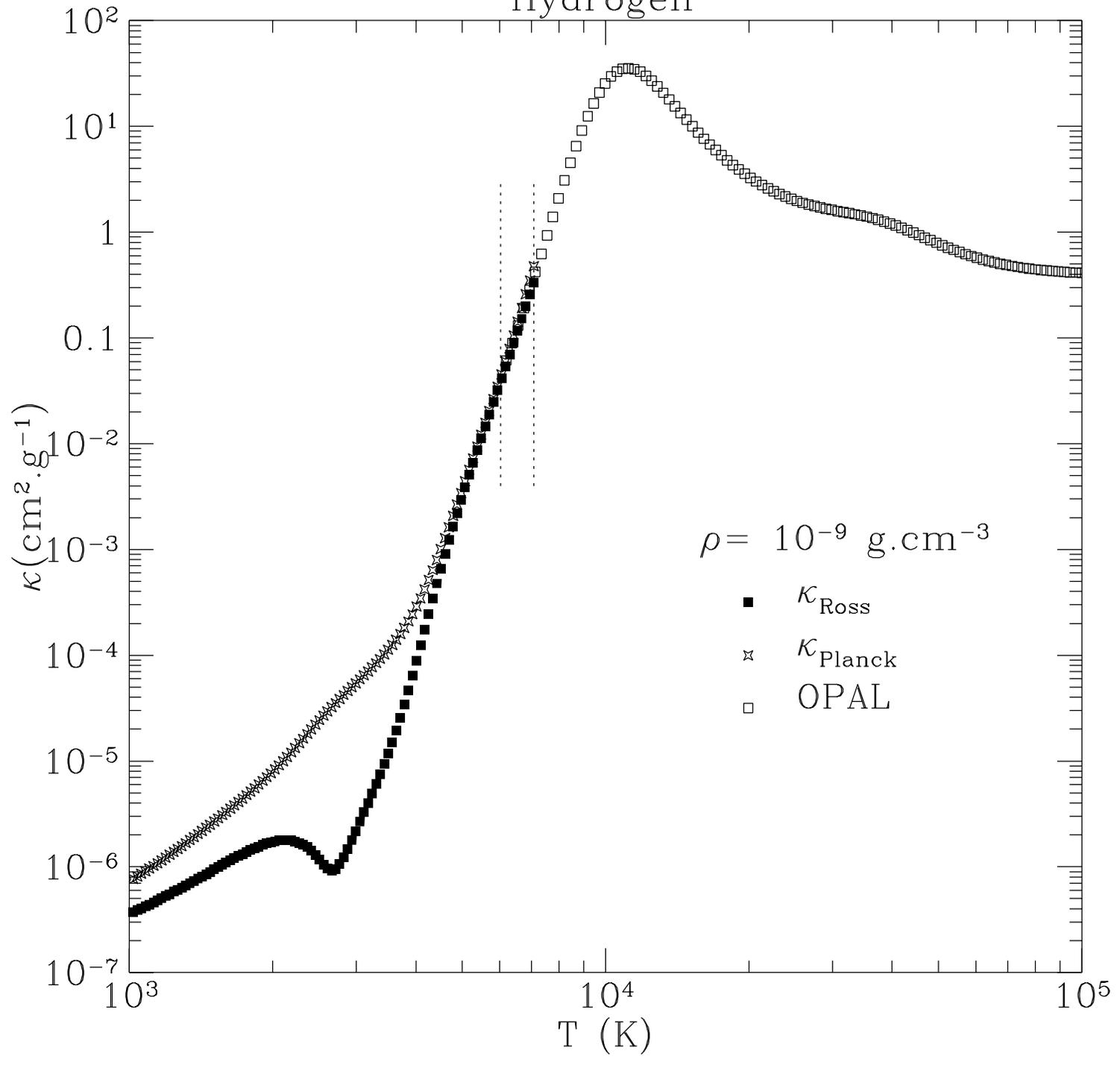
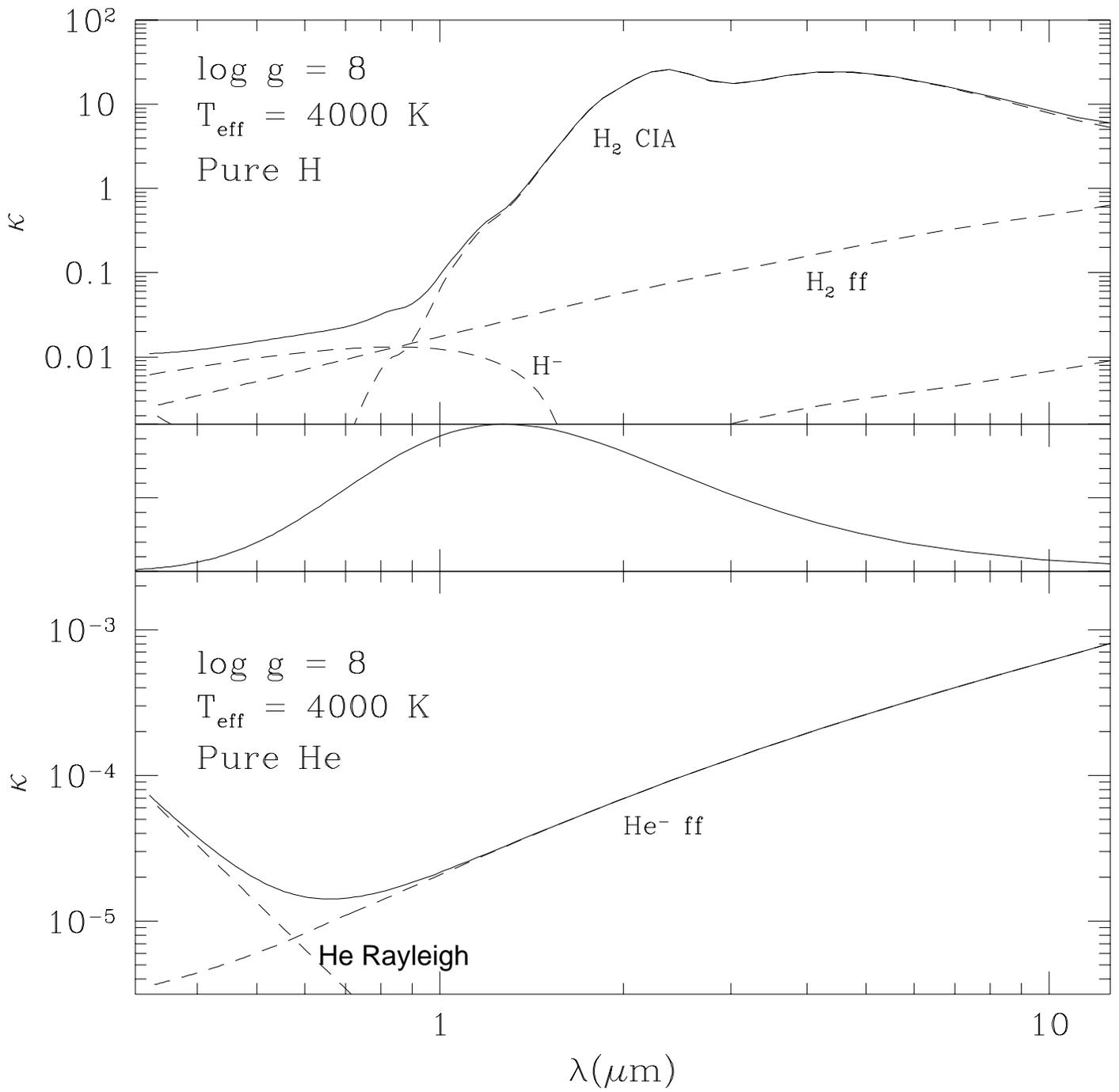
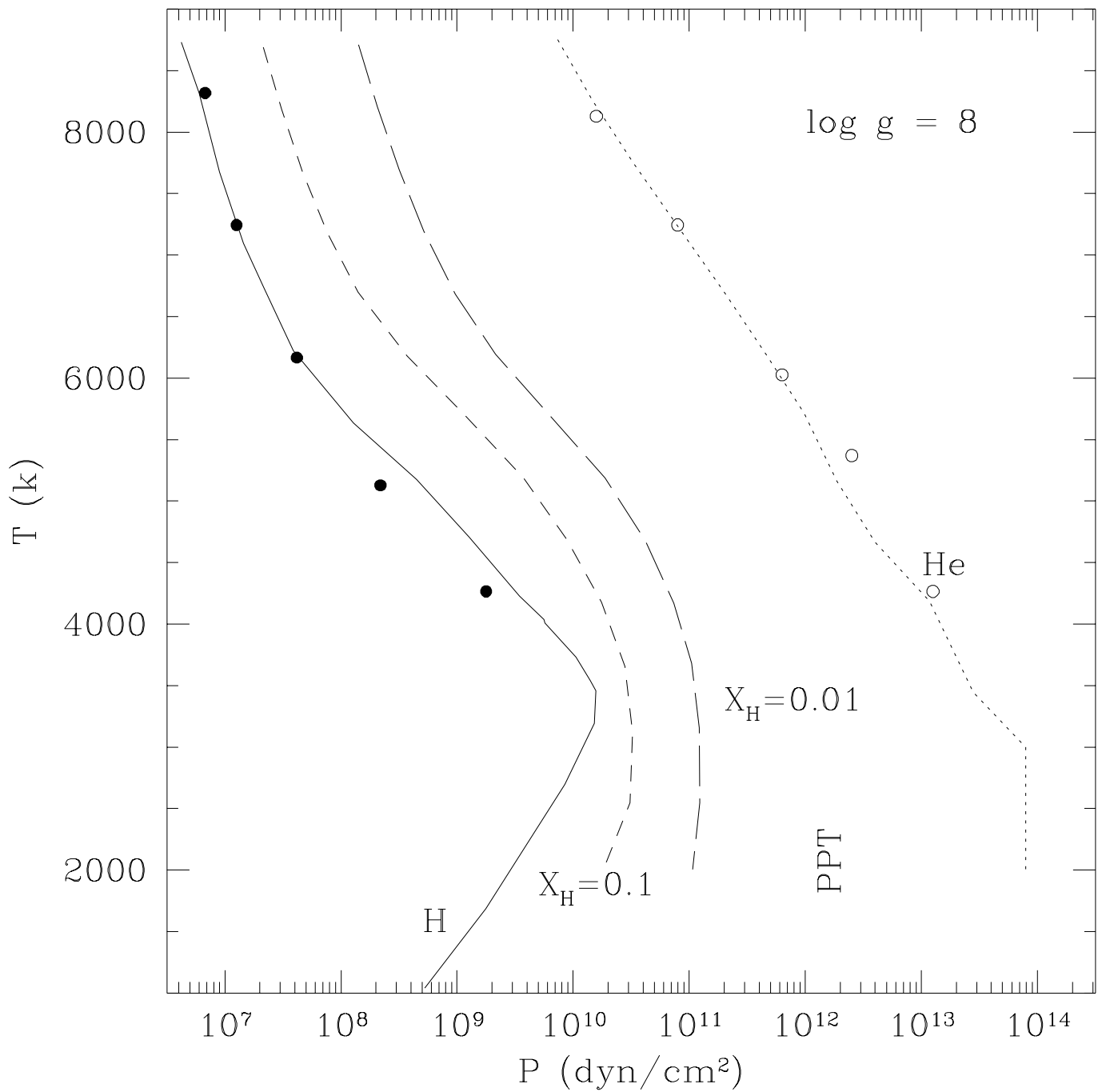
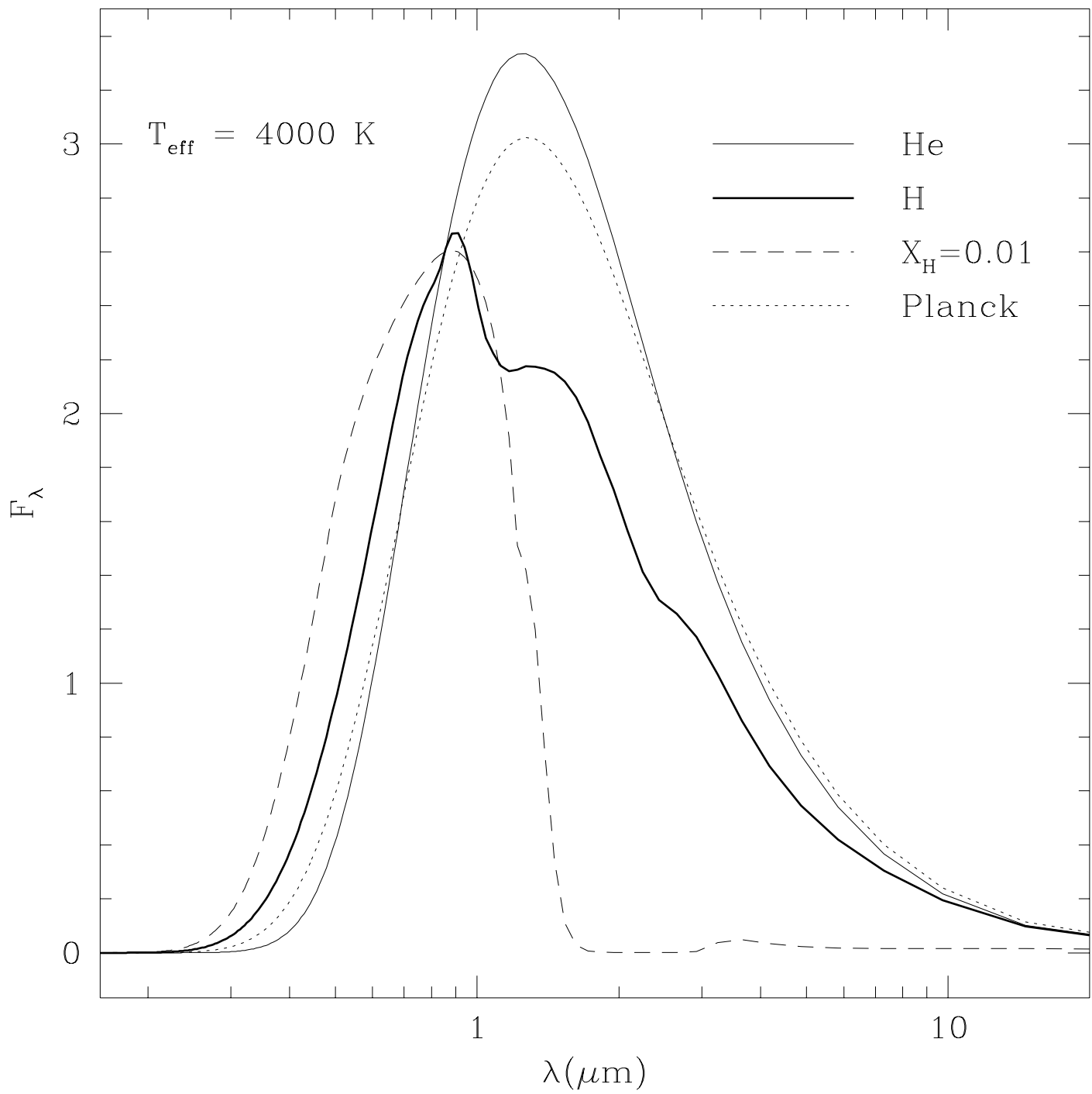
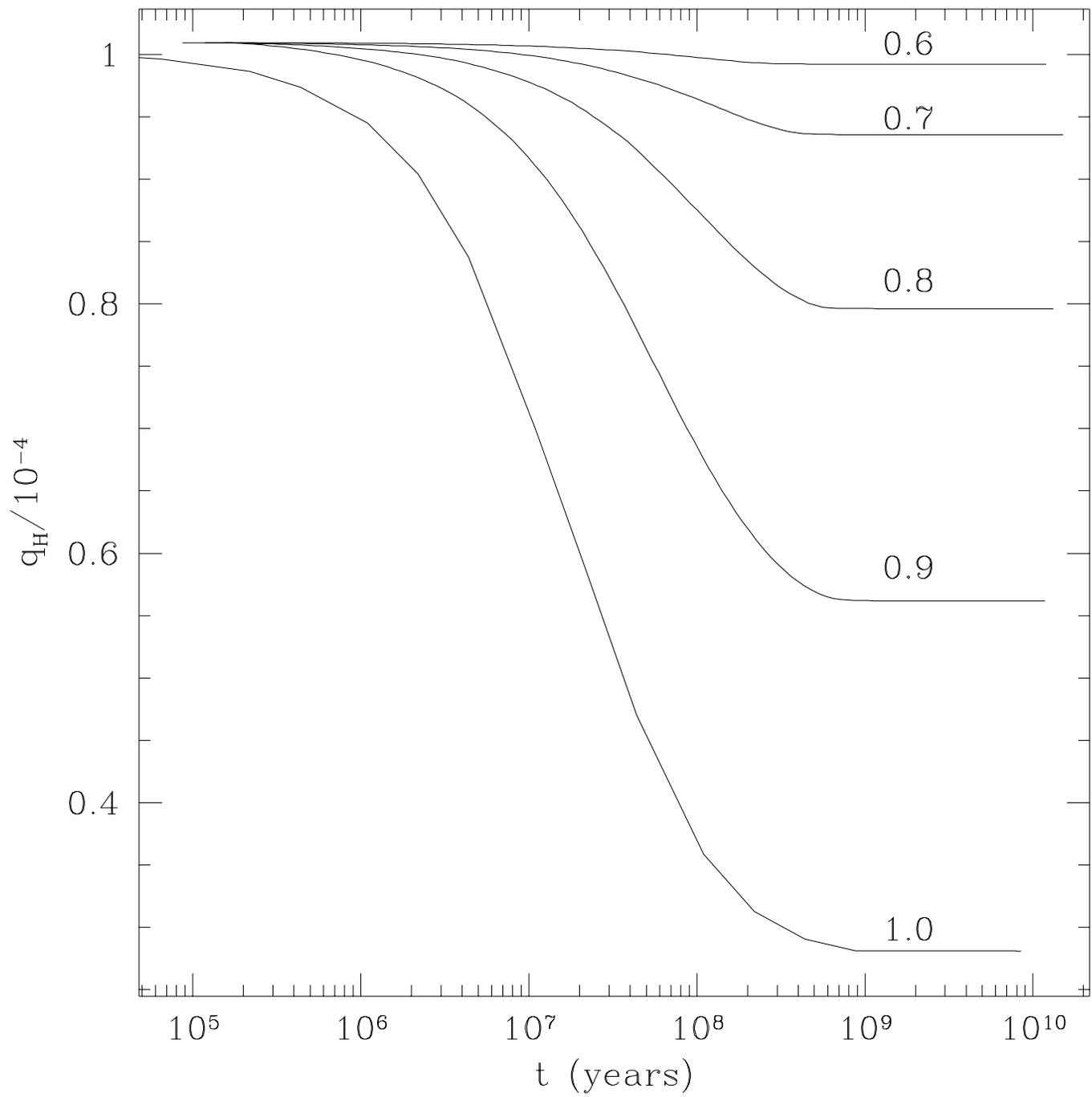
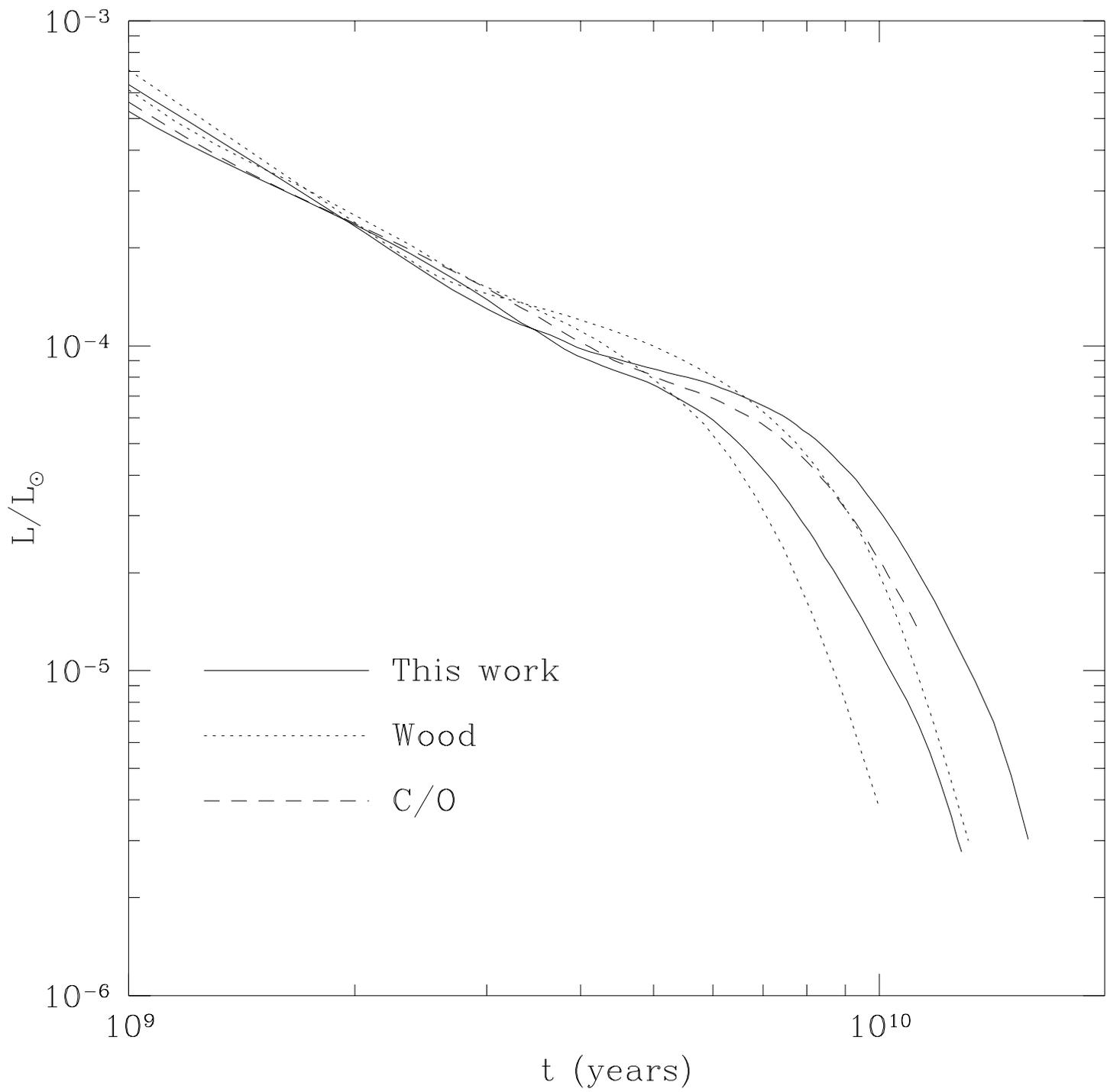
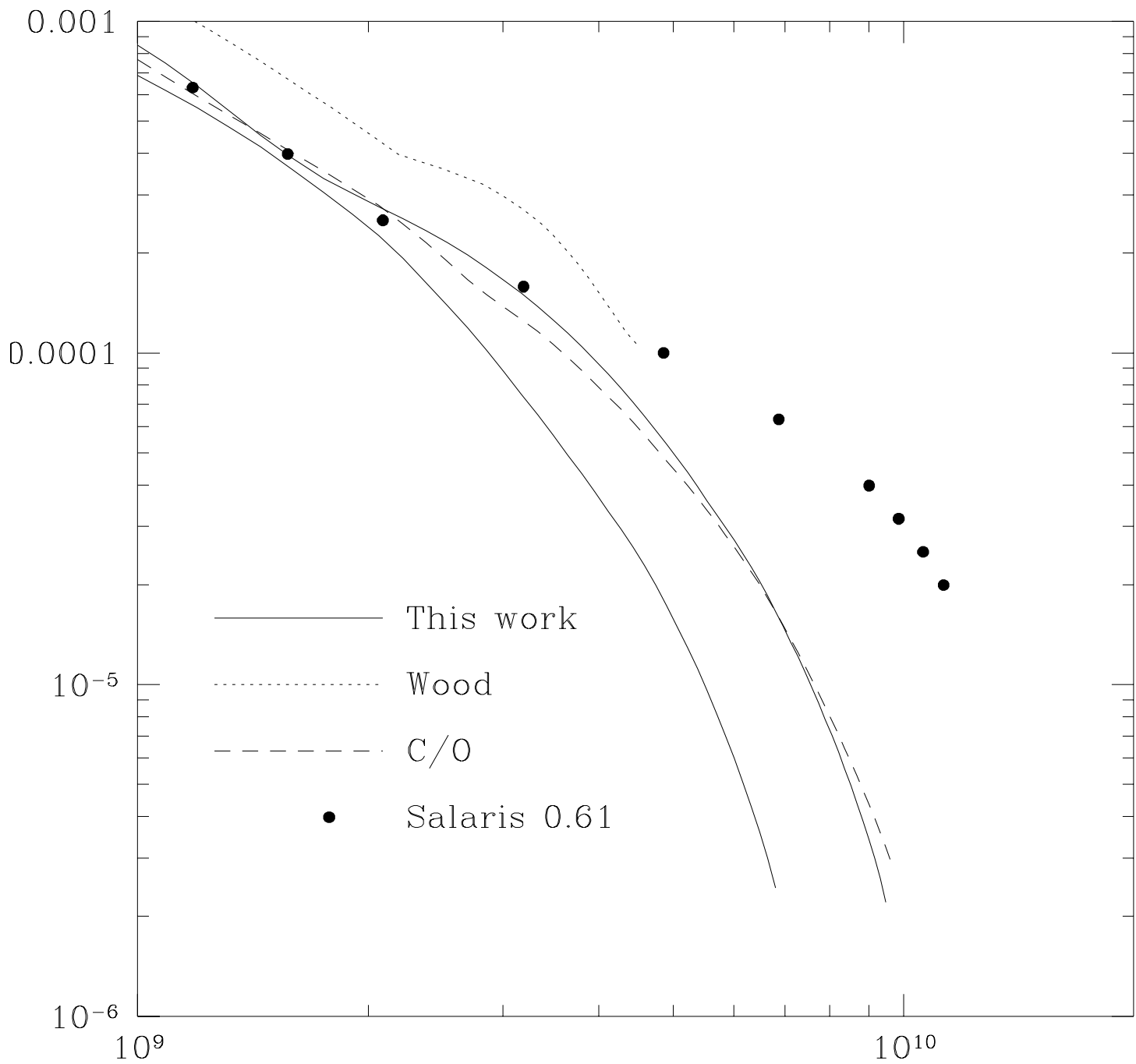
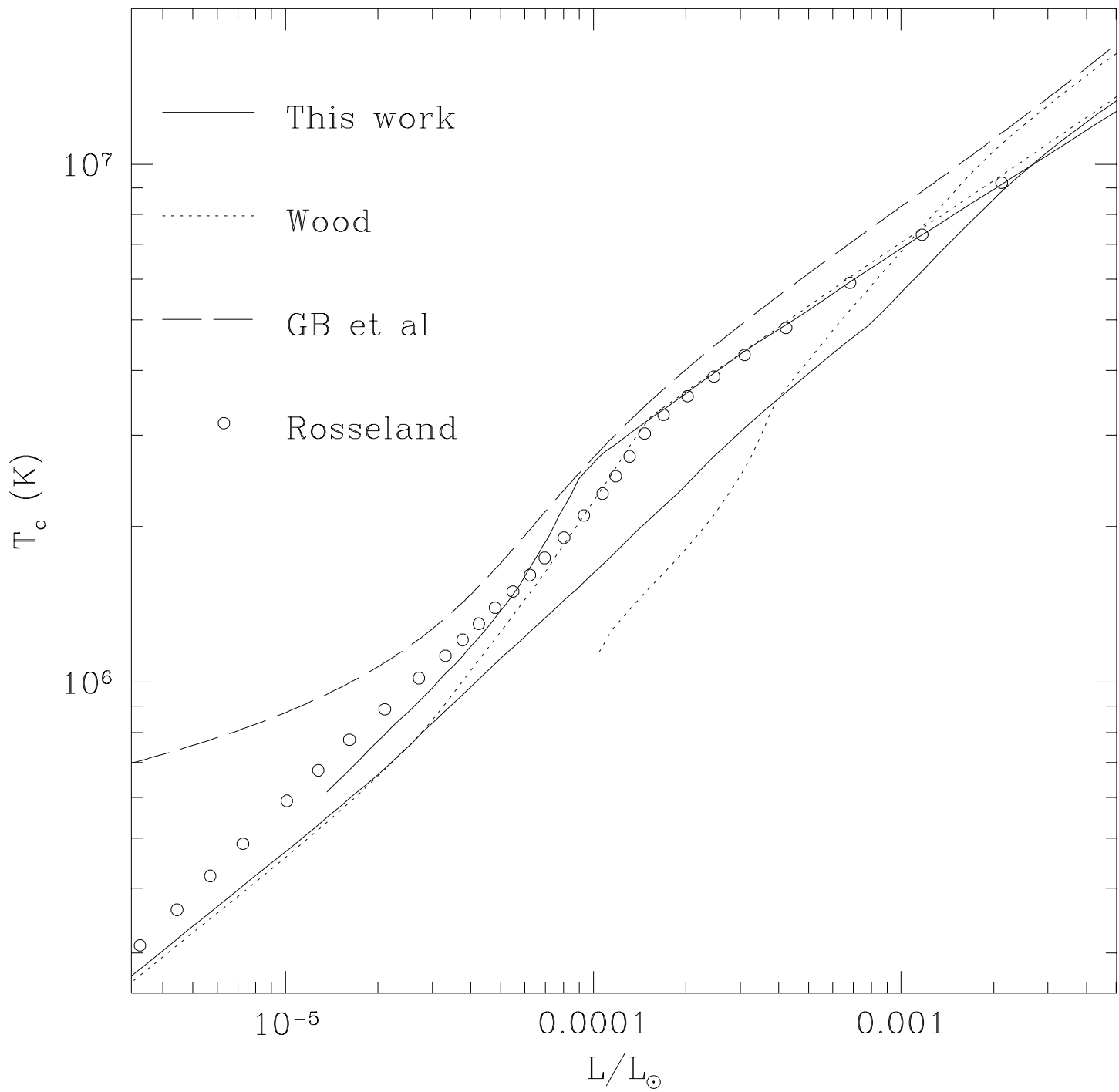
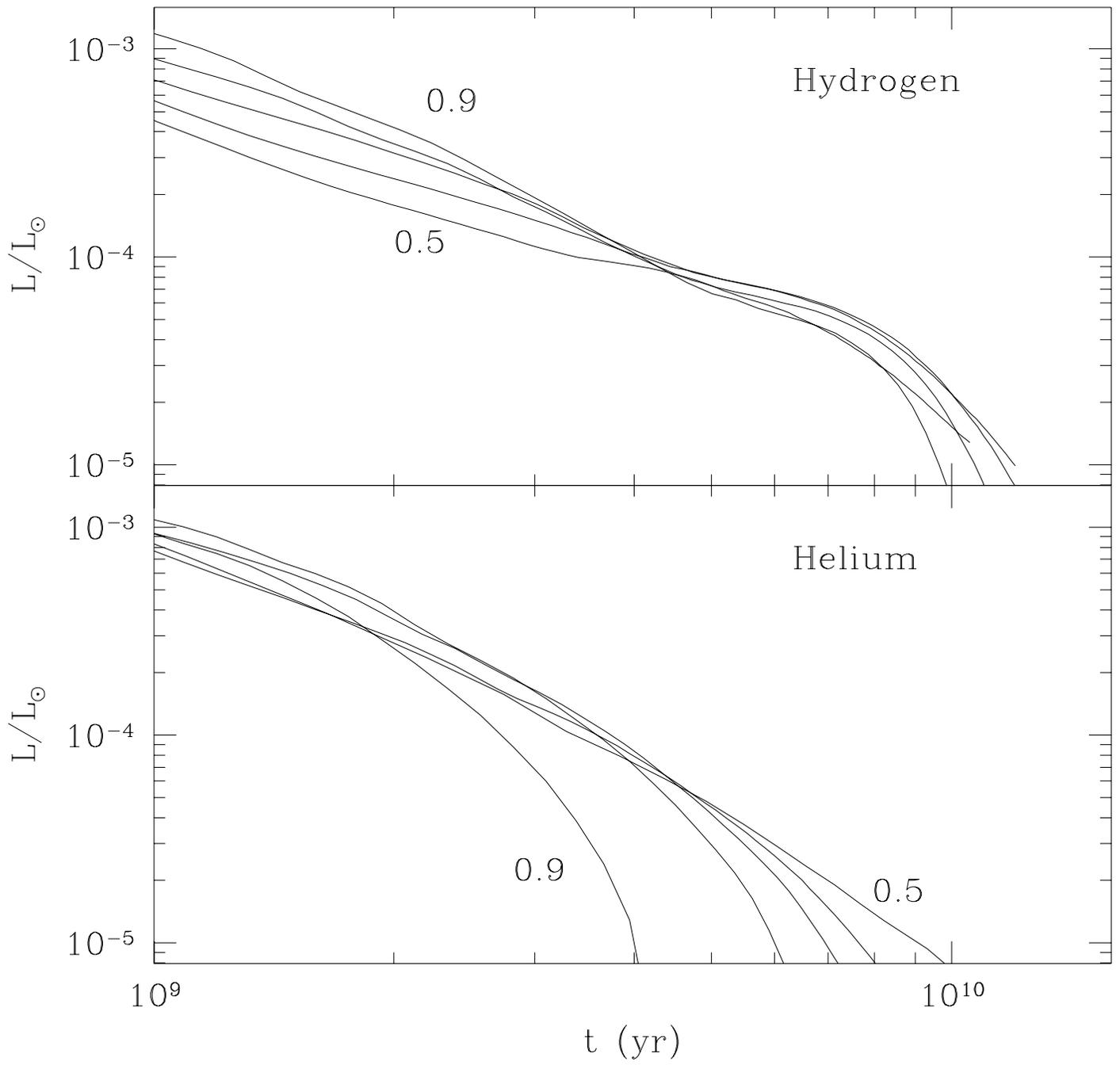
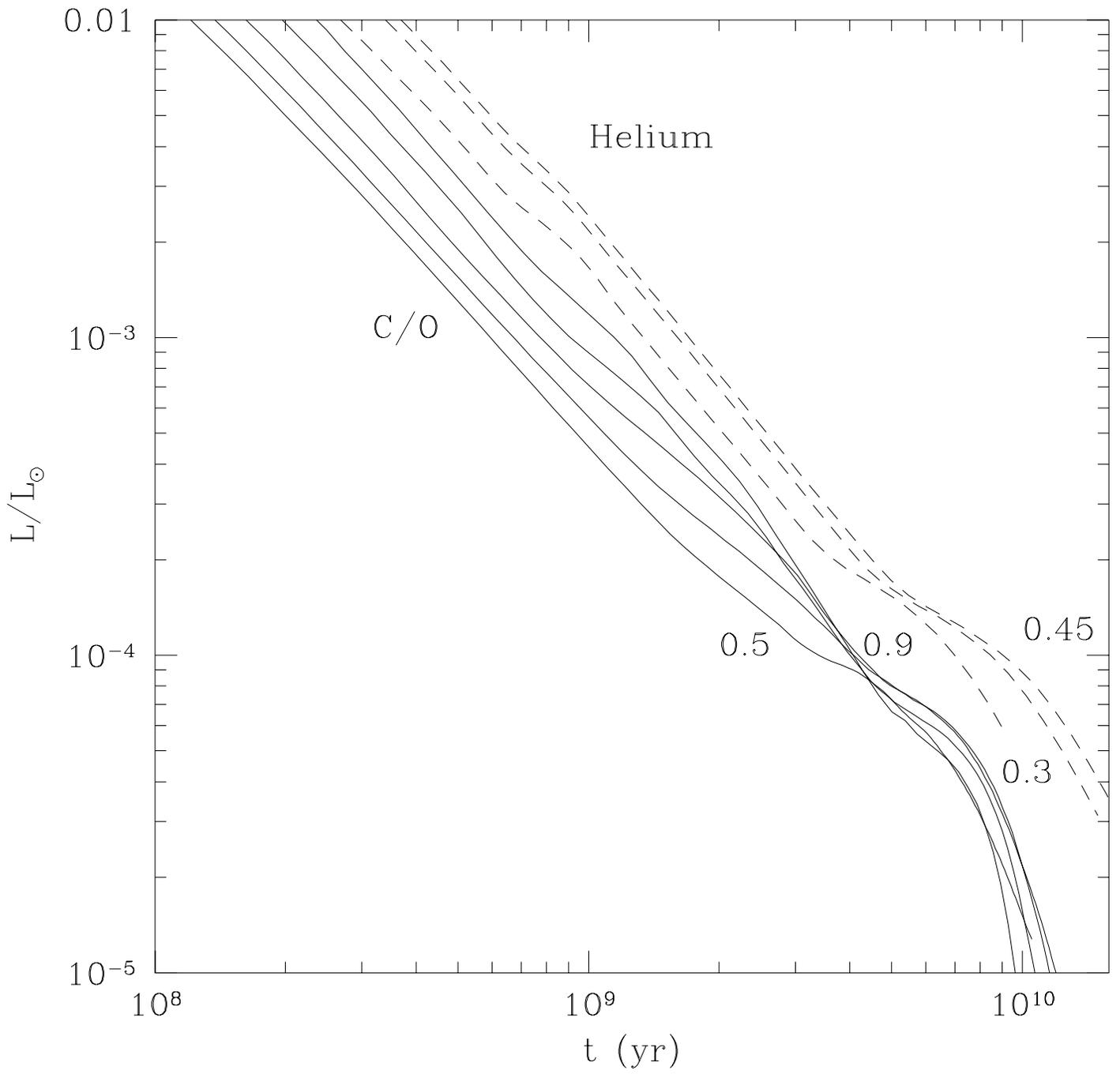
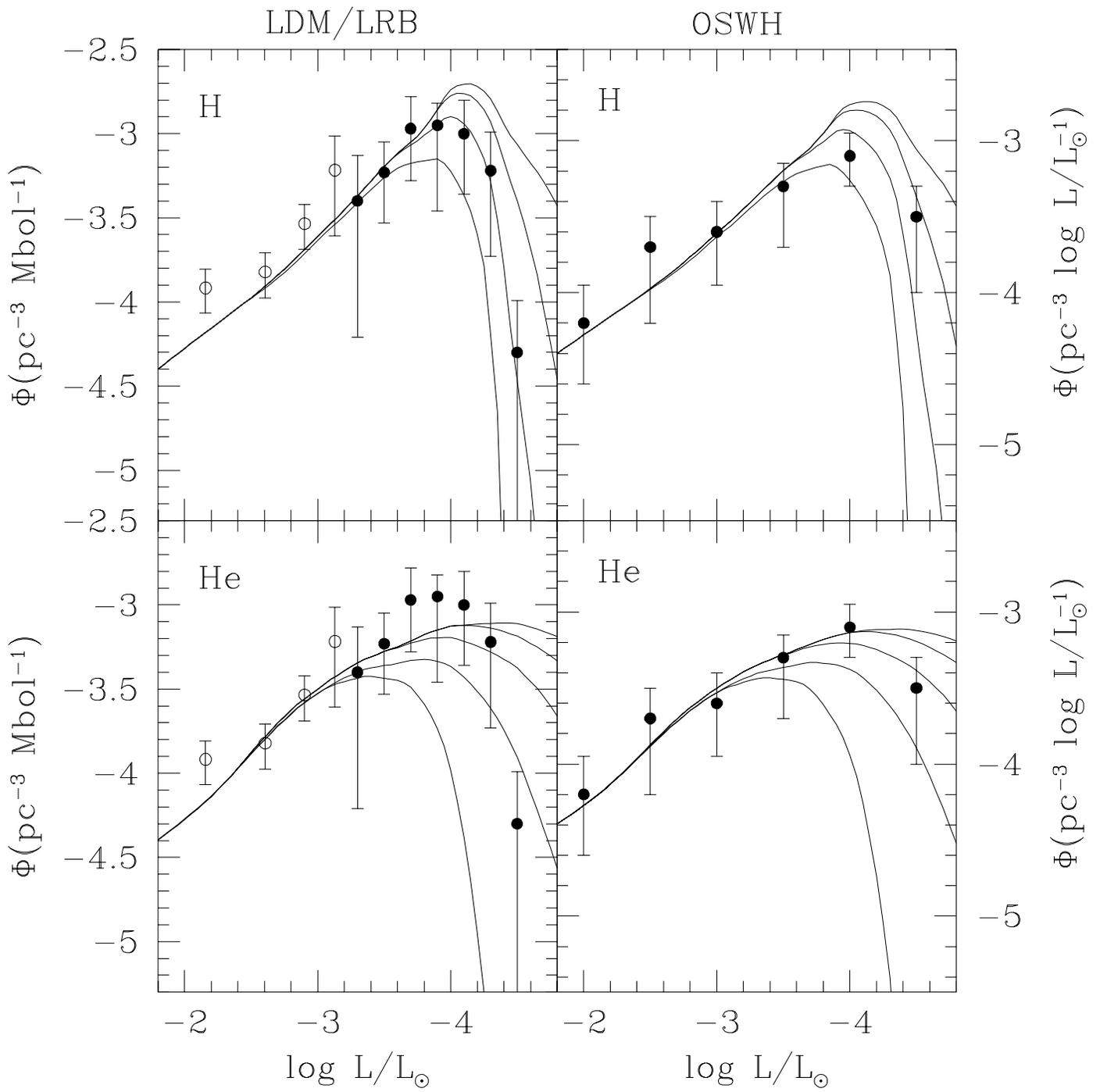
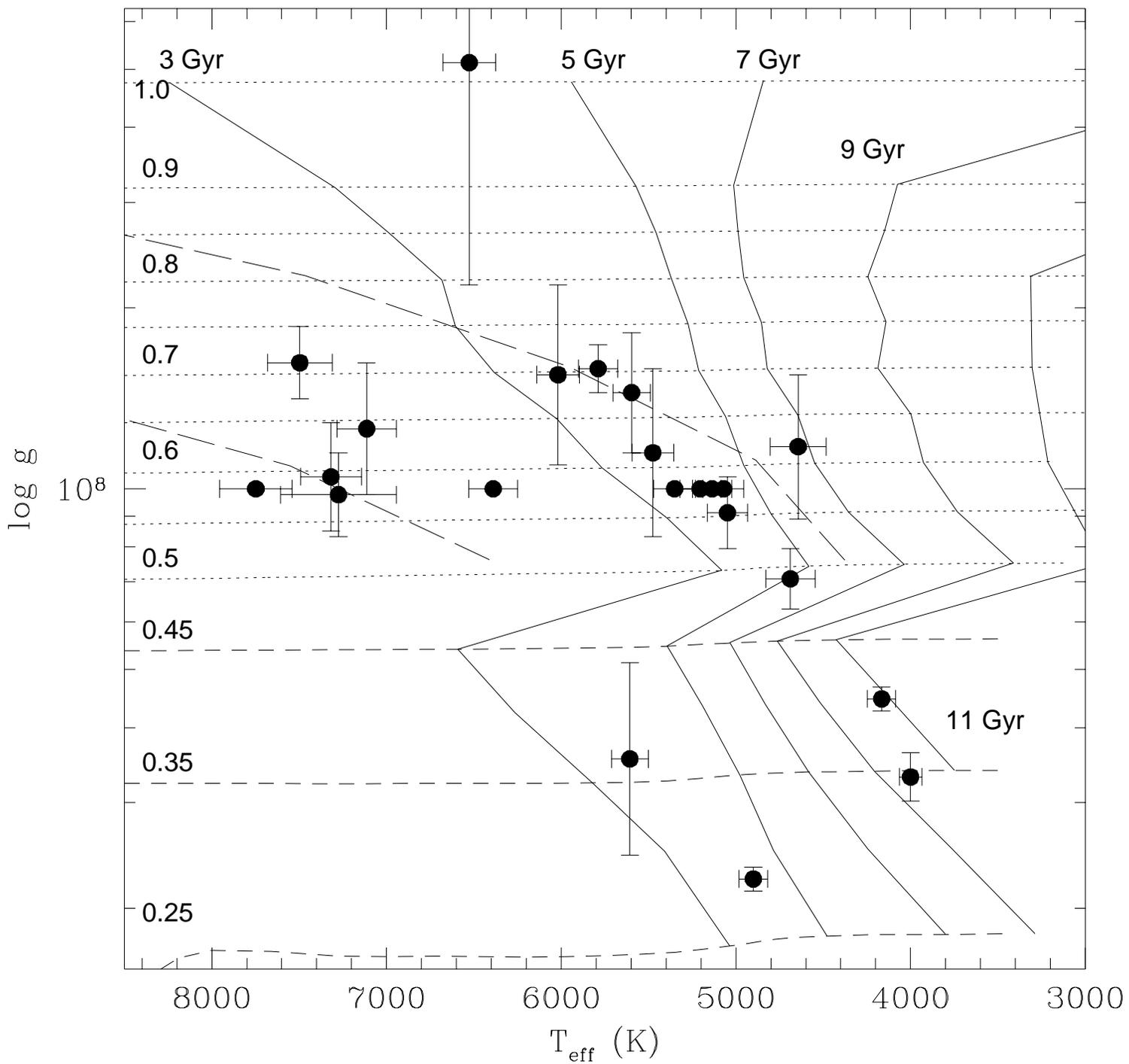
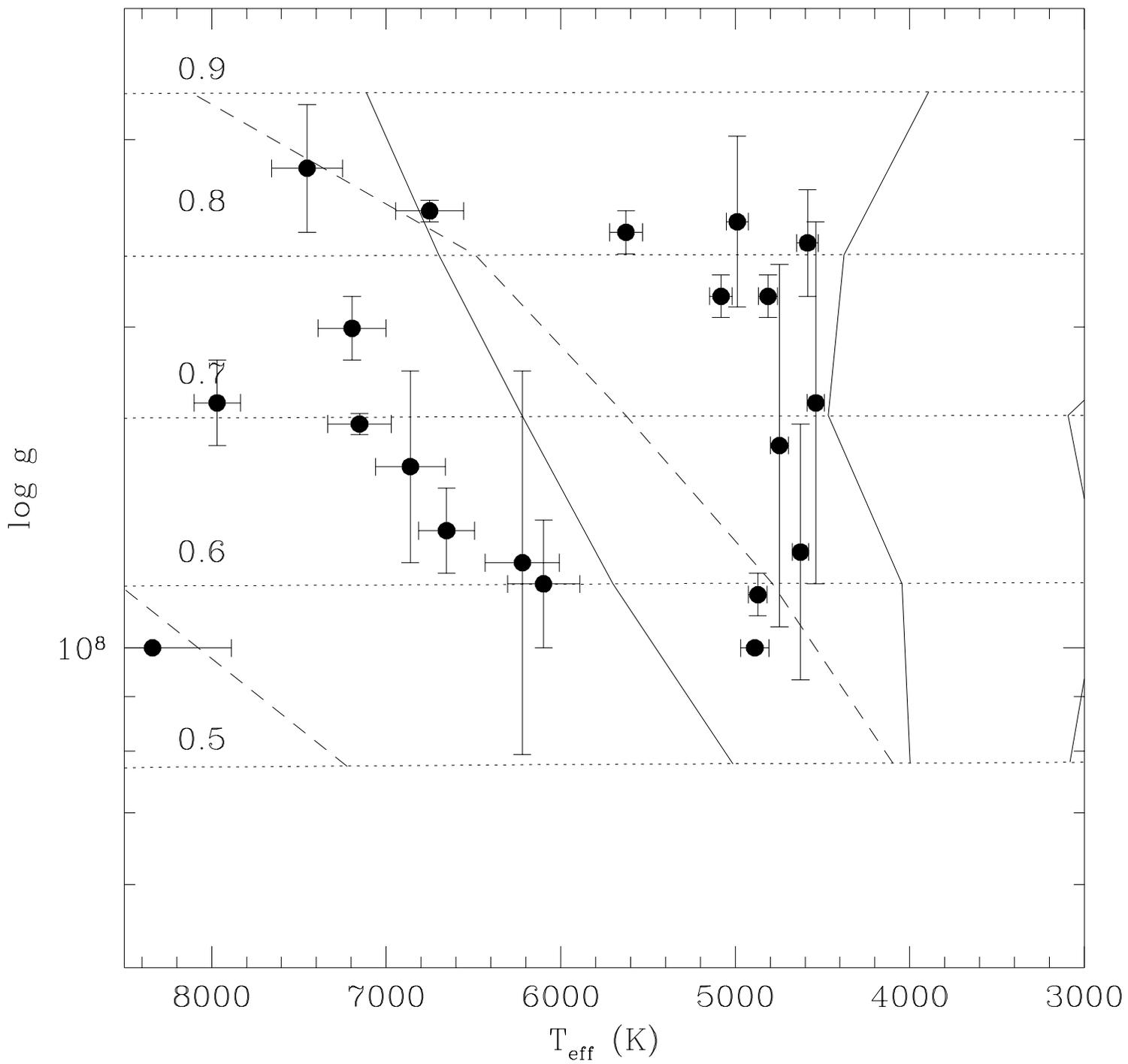
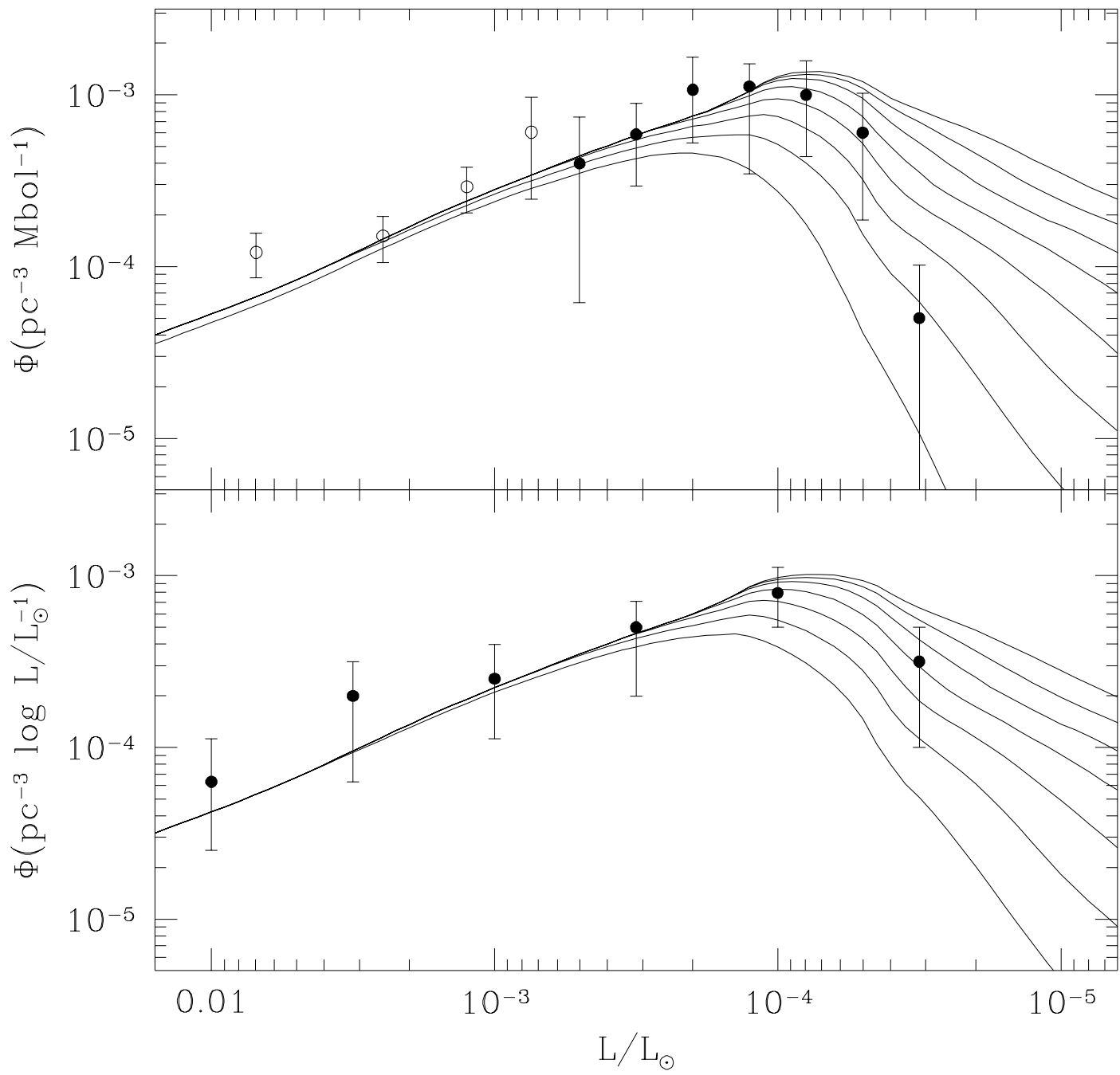
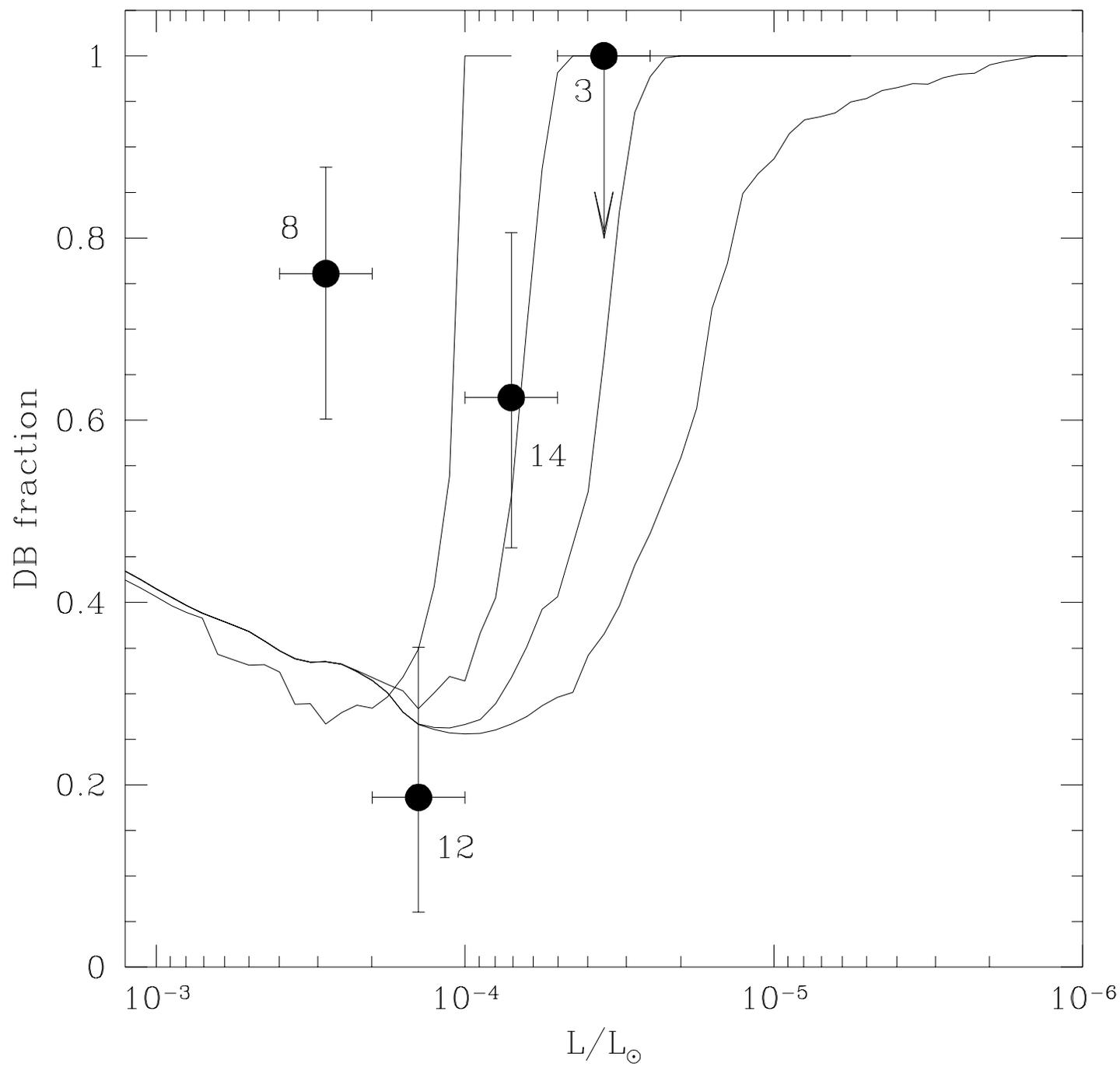
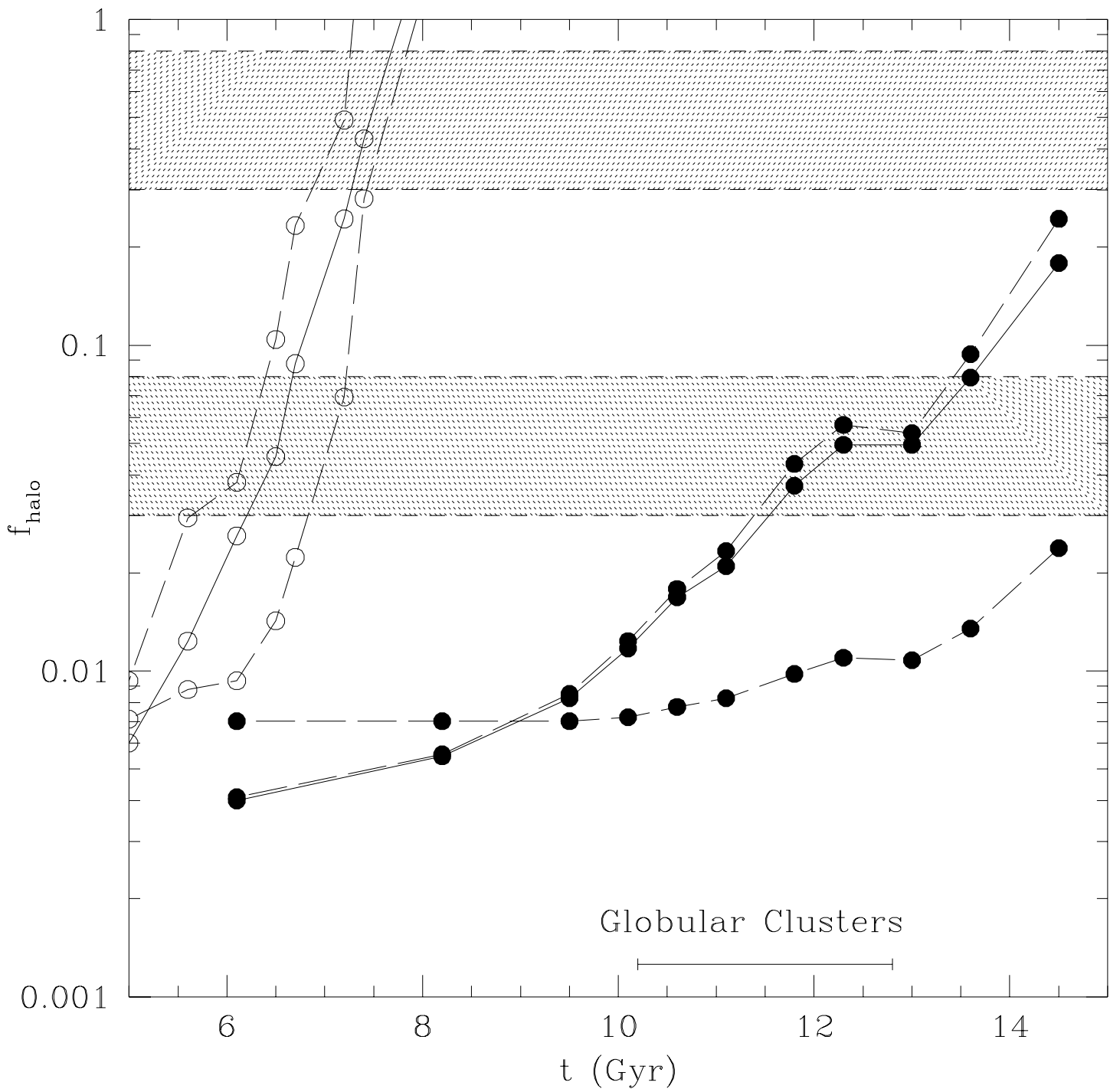
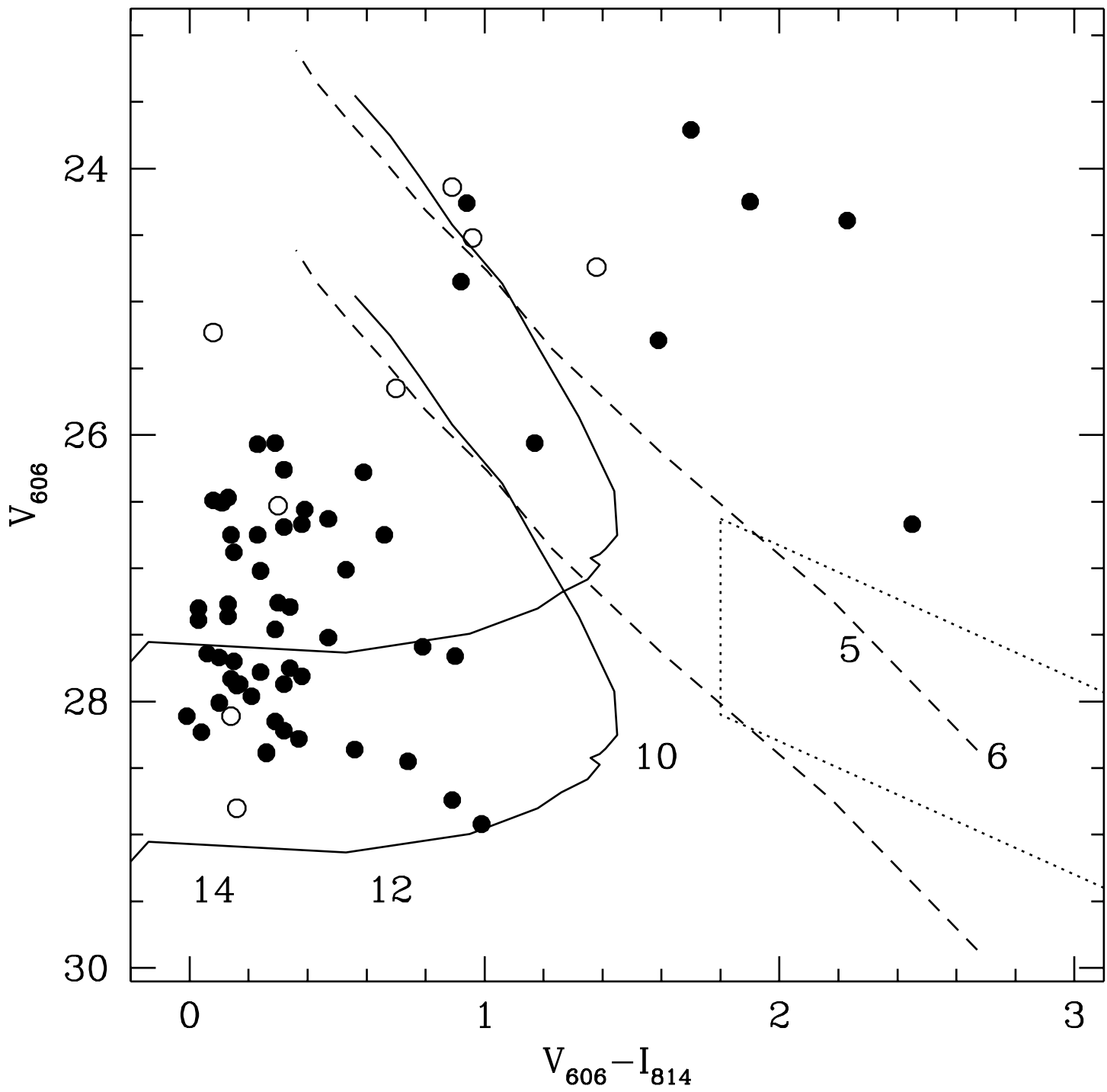
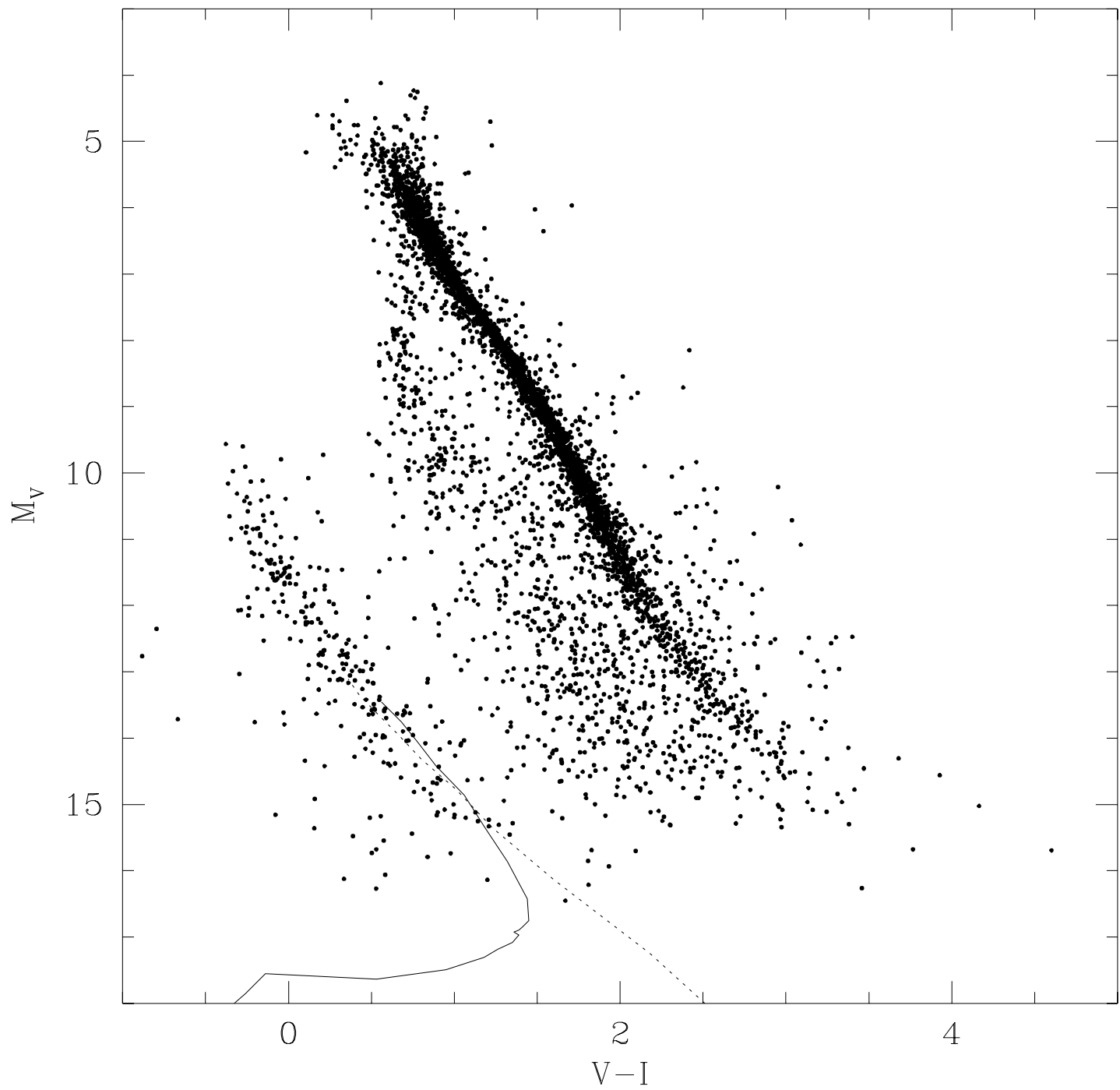
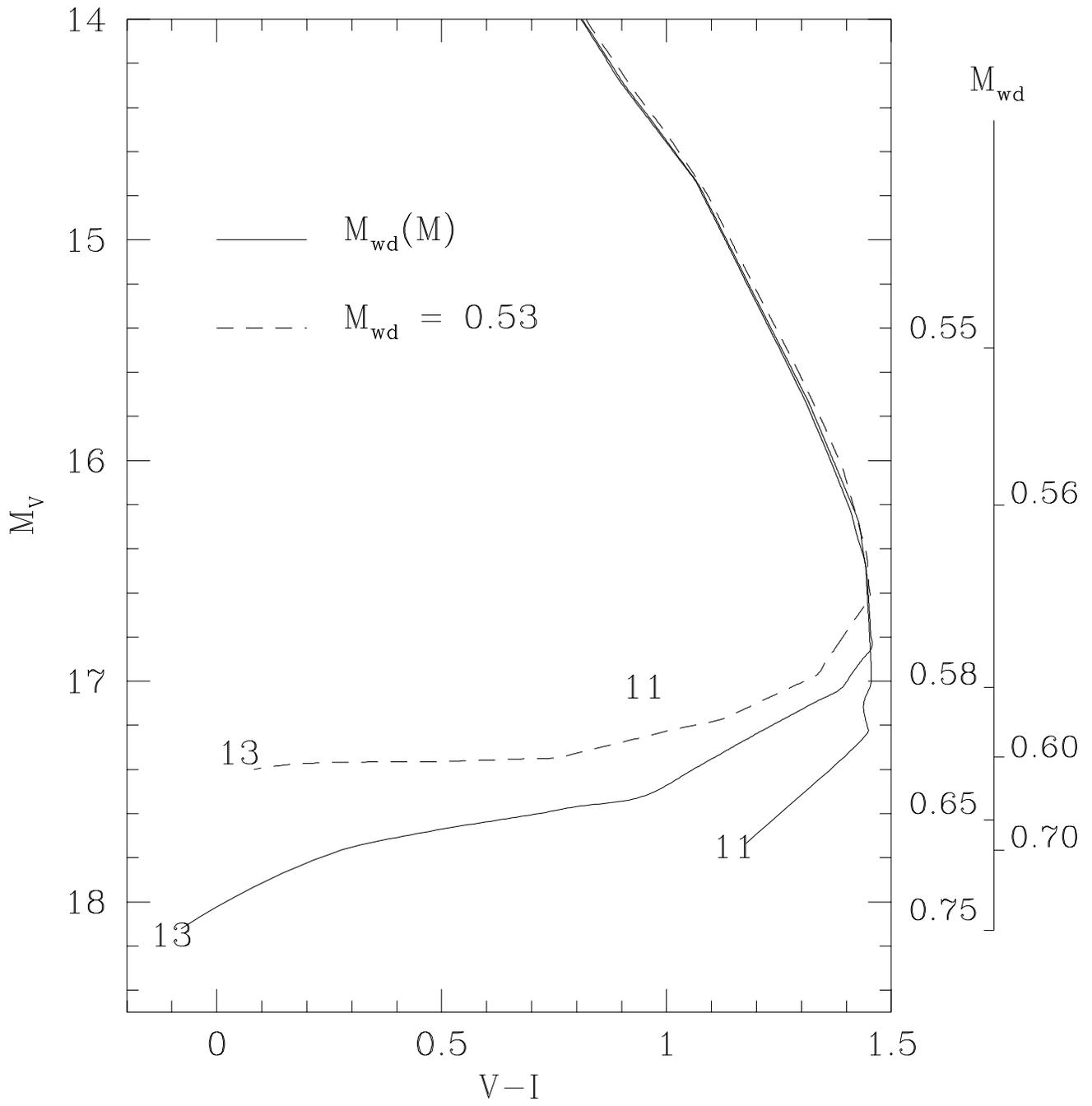
\end{document}